\documentclass[preprint,showpacs,preprintnumbers,unsortedaddress,%
floatfix,amsmath,pra]{revtex4}
\usepackage{bm}
\usepackage{graphicx}%
\usepackage{calc}%
\usepackage{amsfonts}%
\usepackage{times}
\usepackage{dcolumn}
\begin{document}
\title{
Direct mapping between exchange potentials  of Hartree-Fock and
Kohn-Sham schemes as origin  of orbitals proximity
}
\author{ M. Cinal}
\affiliation{Institute of Physical Chemistry of the Polish Academy
of Sciences, ul. Kasprzaka 44/52, 01--224 Warszawa, Poland}
\date{\today}
\begin{abstract}

It is found
that, in closed-$l$-shell atoms, the exact  local exchange potential $v_{\text{x}}(\bf r)$ of
the density functional theory (DFT)
is very well represented, within the region of every atomic shell, by each of the
suitably shifted potentials obtained
with the non-local Fock exchange operator for  the individual Hartree-Fock (HF) orbitals
belonging to this shell. Consequently,
the continuous piecewise  function built of
shell-specific exchange potentials, each defined as the weighted average of the shifted
orbital exchange potentials corresponding 
to a given shell, yields
another highly-accurate representation of $v_{\text{x}}(\bf r)$.
These newly revealed properties are {\em not} related to the well-known step-like shell structure
in the response part of $v_{\text{x}}(\bf r)$,
but they result from specific relations satisfied by the HF  orbital exchange potentials.
These relations explain the outstanding proximity of the occupied Kohn-Sham and HF orbitals
as well as  the high quality of the Krieger-Li-Iafrate
 and  localized HF (or, equivalently, common-energy-denominator)
approximations to the DFT exchange potential $v_{\text{x}}(\bf r)$.
The constant shifts added to the HF orbital exchange potentials, to map them onto $v_{\text{x}}(\bf r)$,
are nearly equal to the differences between the energies of the corresponding  KS and HF orbitals.
It is discussed why these differences are positive and grow when
the respective orbital energies become lower for inner orbitals.

\end{abstract}
\maketitle
\section{INTRODUCTION}
\label{sec-intro}
Representing the quantum state of a many-electron system in terms of
one-electron orbitals is simple and theoretically attractive
approach. Such description is realized in the Hartree-Fock (HF)
method \cite{johnson07}, as well as in the Kohn-Sham (KS) scheme of
the density-functional theory (DFT)  \cite{PY89,DG91,FN03}.
The latter is  an efficient and robust tool which is
now routinely applied in the calculations of electronic properties of
molecules, even very large and complex, and condensed-matter structures.
Though the
KS scheme is formally accurate, the one-body KS potential contains
the exchange-correlation (xc) potential $v_{\text{xc}}$, whose exact dependence on the
electron density remains unknown. It is usually treated within the
local-density or generalized-gradient
approximations (LDA, GGA),
despite the well-known shortcomings of the LDA and GGA xc potentials
(especially the self-interaction errors). Some of these
deficiencies are removed when the exact form (in terms of the occupied KS orbitals)
is used for the exchange part $E_{\text{x}}$ of the xc energy.
The exact
exchange potential  $v_{\text{x}}$ is then found from $E_{\text{x}}$ by means of the integral
equation resulting from the optimized-effective-potential (OEP)
approach [\onlinecite{KLI92}(a),\onlinecite{EV93,GKKG00,E03,KK08}] or by using the
recently developed method based on the differential equations for
the orbital shifts \cite{KP03,CH07}; another method based on the direct
energy minimization  with respect to the KS-OEP potential (expressed in a finite basis)
\cite{YW02} suffers from  convergence problems \cite{SSD06}
 which are not fully resolved yet and they are still under study\cite{HBBY07,HBY08}.
The exact potential $v_{\text{x}}$  is 
free from self-interaction and it has correct asymptotic dependence
($-1/r$ for finite systems) at large distances $r$  from the system;
thus, unlike the HF, LDA or GGA potentials, it produces correct unoccupied states.
In the DFT, the approximation, in which the exchange is
included exactly but the correlation energy and potential are
neglected, is known as the exchange-only KS scheme --- it is applied
in the present investigation.
The full potential $v_{\text{xc}}$ can also be found by means of the OEP approach
when the DFT total energy includes, besides the exact $E_{\text{x}}$,
the correlation energy $E_{\text{c}}$ depending on all (occupied
and unoccupied) KS orbitals and orbital energies \cite{E03}.
This makes such computation tedious, to a level undesirable in the DFT,
since it involves calculating $E_{\text{c}}$ with  the quantum-chemistry
methods, like the M{\o}ller-Plesset many-body perturbation approach.

Defined to yield the true electron density,
the KS one-electron orbitals  have no other direct physical meaning
since they formally refer to a {\em  fictitious} system of {\em non-interacting}
electrons. However, it is a common practice to use these orbitals in
calculations of various electronic properties; in doing so the
$N$-electron ground-state wave function $\Psi_0$ of the physical
(interacting) system is approximated with the single determinant
built of the KS orbitals. This approximate approach is justified by
(usually) sufficient accuracy of the calculated quantities, which is
close to, or often better than, that of the HF results \cite{Bour00}.
It seems that the
success of the DFT calculations would not be possible if the  KS
determinant, though being formally non-physical, was not close to
the HF determinant which, outside the DFT,   is routinely
used to approximate the wave function $\Psi_0$  of
the {\em real} system.  Therefore, understanding this proximity is
certainly very important for the fundamentals of the DFT.

Previous calculations [\onlinecite{KLI92}(a),\onlinecite{ZP92_93,Nagy93,CES94}]
have shown that,
not only the whole KS and HF determinants \cite{Bour00,DSG01} and the corresponding electron densities
[\onlinecite{KLI92}(a),\onlinecite{GE95, SMVG97,SMMVG99}], but also
the {\em individual  occupied} KS and HF orbitals,
$\phi_{a\sigma}({\bf r})$ and $\phi_{a\sigma}^{\text{HF}}({\bf r})$, in atoms
are so close to each other  that
they are virtually indistinguishable
(here the orbitals, dependent on the electron position $\bf r$
and the spin $\sigma=\downarrow$,  $\uparrow$,
are numbered with index $a=1,\ldots,N_{\sigma}$; $N_{\downarrow}+N_{\uparrow}=N$).
This property  is particularly remarkable for the exchange-only KS orbitals
which differ so minutely from  the HF orbitals
that, for atoms,  the OEP total energy 
is only several mhartrees higher than the HF energy $E_{\text{HF}}$
\cite{KLI92,EV93,KK08,GE95}.
The outstanding proximity of the KS and HF orbitals is surprising in view  of the
obvious difference between the exchange operators
in the KS and HF one-electron hamiltonians (see below) and the fact that
the corresponding KS and HF atomic orbital energies,
$\epsilon_{a\sigma}$ and $\epsilon_{a\sigma}^{\text{HF}}$,
differ substantially, up to several hartrees for core orbitals in atoms
like Ar, Cu \cite{EV93,KK08} [except for the KS and HF energies of
the highest-occupied molecular orbital (HOMO) which are almost identical].
This apparent contradiction has not yet been resolved;
in Ref. \cite{IL03}  it is suggested
that the KS and HF determinants are close to each other
``since the kinetic energy is much
greater than the magnitude of the exchange energy".

The present paper investigates the proximity of the KS and HF orbitals
and it reveals that, in closed-$l$-shell
atoms, there exists a direct mapping  between
the HF orbital local exchange potentials
$v_{\text{x}a\sigma}^{\text{HF}}({\bf r})$
and the DFT exact local exchange potential
$v_{\text{x}\sigma}({\bf r})$.
The former are
specific to each HF orbital $\phi_{a\sigma}^{\text{HF}}({\bf r})$ and are defined as
\begin{equation}
\label{eq-vxa-hf}
v_{\text{x}a\sigma}^{\text{HF}}({\bf r}) \equiv
\frac{\hat{v}_{\text{x}\sigma}^{\text{F}}\phi_{a\sigma}^{\text{HF}}({\bf r}) }
{\phi_{a\sigma}^{\text{HF}}({\bf r}) }\;
\end{equation}
with the Fock exchange non-local operator  $\hat{v}_{\text{x}\sigma}^{\text{F}}({\bf r}) $
within the HF approximation that describes
the interacting system.
The DFT exchange potential  $v_{\text{x}\sigma}({\bf r})$
is common for all orbitals relevant to the KS
non-interacting $\sigma$ subsystem.
This potential  is found to be very well represented, within the region
of each atomic shell, by the {\em individual}, suitably shifted potentials
$\tilde{v}_{\text{x}a\sigma}^{\text{HF}}({\bf r}) =v_{\text{x}a\sigma}^{\text{HF}}({\bf r})+C_{a\sigma}$
obtained for the HF orbitals that belong to this shell;
the constant shifts $C_{a\sigma}$ are orbital-specific.
As a result, for each shell, the weighted average of the
potentials $\tilde{v}_{\text{x}a\sigma}^{\text{HF}}({\bf r})$
corresponding to the orbitals from this shell yields
the shell-specific exchange potential that also represents $v_{\text{x}\sigma}({\bf r})$
with high accuracy within the shell region.
The revealed mapping between $\tilde{v}_{\text{x}a\sigma}^{\text{HF}}({\bf r})$
and $v_{\text{x}\sigma}({\bf r})$ is shown
to have origins in the specific relations satisfied by
the HF orbital exchange potentials. Thus, the
proximity of the KS and HF orbitals is explained.
Simultaneously, it becomes
clear why,  in atoms, the exact exchange potential
$v_{\text{x}\sigma}(r)$  (where $r=|{\bf r}|$) has the characteristic structure of a
piecewise function where each part spans over the region of an
atomic shell and it has distinctively different slope
$dv_{\text{x}\sigma}(r)/dr$  in consecutive shells \cite{leeuwen96}.

%
%
%
The specific properties
of $v_{\text{x}a\sigma}^{\text{HF}}({\bf r})$ are also shown to be directly
responsible for
the high quality of
the  approximate representations of
the exact exchange potential $v_{\text{x}\sigma}({\bf r})$
that are obtained in  the Krieger-\-Li-Iafrate(KLI) \cite{KLI92}  and
 localized HF (LHF)\cite{DSG01} approximations,
the latter of which is equivalent to the
common-energy-denominator approximation (CEDA) \cite{GB01}.
The constant shifts $C_{a\sigma}$,
needed  to map the HF potentials
$v_{\text{x}a\sigma}^{\text{HF}}({\bf r})$ onto $v_{\text{x}\sigma}({\bf r})$,
are shown to be nearly equal to
$\epsilon_{a\sigma}-\epsilon_{a\sigma}^{\text{HF}}$.
This leads to better understanding
why, for each KS occupied orbital (other than the HOMO), its energy $\epsilon_{a\sigma}$
is  higher than the corresponding HF energy $\epsilon_{a\sigma}^{\text{HF}}$
and the difference between these two energies is larger for the core orbitals
than for the valence ones.
Finally, it is shortly argued that
the presently revealed properties of the KS and HF exchange potentials
do {\em not} result from
the well-known step-like shell structure present
in the response part $v_{\text{x}\sigma}^{\text{resp}}({\bf r})$ of
the exchange potential  \cite{leeuwen94,leeuwen95}.




\section{THEORY}
\label{sec-theory}

\subsection{Hartree-Fock method and optimized-effective-potential approach}
\label{sec-theory-HF-OEP}

The HF one-electron spin-orbitals $\phi_{a\sigma}^{\text{HF}}(\bf
r)$ are obtained by minimizing the mean value $\langle \Psi |\hat
H|\Psi \rangle$ where $\hat H$ is the Hamiltonian of the
$N$-electron interacting system and $\Psi $ belongs to the subspace
$\Omega_N^{\text{det}}$
of normalized $N$-electron wave functions that are single
Slater determinants built of one-electron  orbitals. Similar  minimization
is carried out in the exchange-only OEP method, but there is the additional
constraint that for every trial
determinant all $N_{\sigma}$ constituent spin-orbitals
$\phi_{a\sigma}({\bf r})$ satisfy the KS  equation with some local KS
potential $v_{\text{s}\sigma}({\bf r})$. The  minimizing potential
$v_{\text{s}\sigma}({\bf r})=v_{\text{s}\sigma}^{\text{OEP}}({\bf r})$,
yields, after subtracting from it the external
$v_{\text{ext}}({\bf r})$ and electrostatic $v_{\text{es}}({\bf r})$
terms, the exact  exchange potential
$v_{\text{x}\sigma}({\bf r})= v_{\text{x}\sigma}^{\text{OEP}}({\bf r})$
(corresponding to the density $n_{\sigma}$ calculated from occupied $\phi_{a\sigma}$),
so that we have
\begin{equation}
\label{eq-vs-ks}
v_{\text{s}\sigma}({\bf r})=v_{\text{ext}}({\bf r})+v_{\text{es}}({\bf r})+v_{\text{x}\sigma}({\bf r}) \, .
\end{equation}
It has to be stressed here that the proximity of the exchange-only KS and HF orbitals
is  {\em not} readily implied by the fact the two sets of orbitals result from the minimization
of the same functional of energy, i.e., $E[\Psi]=\langle \Psi |\hat H|\Psi \rangle$ where
$\Psi \in \Omega_N^{\text{det}}$.
Indeed, for a suitably chosen model Hamiltonian $\hat{H}$,
the corresponding  HF orbitals $\phi_{a\sigma}^{\text{HF}}({\bf r})$ that minimize $E[\Psi]$
might not be well approximated 
by any set of one-electron (KS) orbitals $\phi_{a\sigma}({\bf r})$
that come from a common local potential $v_{\text{s}\sigma}({\bf r})$.
Then, the latter condition, which is 
imposed on the orbitals $\phi_{a\sigma}({\bf r})$ in the OEP minimization,
would be so restrictive that the obtained KS-OEP orbitals would differ significantly from the HF ones.
Thus, it seems that it is the specific form
of the physical Hamiltonian $\hat{H}$ (with Coulombic interactions)
that actually makes the  close representation of the HF orbitals with the KS ones possible.

The exchange-only KS equation,
satisfied by the corresponding (OEP) orbitals $\phi_{a\sigma}({\bf
r})$ and their energies
 $\epsilon_{a\sigma}$, takes the form
\begin{equation}
\label{eq-ks}
\hat{h}_{\text{s}\sigma}({\bf r})  \phi_{a\sigma}({\bf r}) \equiv
    \Bigl[-\frac{1}{2}\bm{\nabla}^2 +v_{\text{ext}}({\bf r})+v_{\text{es}}({\bf r})+
    v_{\text{x}\sigma}({\bf r}) \Bigr]
    \phi_{a\sigma}({\bf r}) = \epsilon_{a\sigma}\phi_{a\sigma}({\bf r})
\end{equation}
(atomic units are used throughout)
where we put $v_{\text{x}\sigma}({\bf r})=v_{\text{x}\sigma}^{\text{OEP}}({\bf r})$
in the OEP case.
The total electron density
$n_{\text{tot}}({\bf r})=n_{\uparrow}({\bf r})+n_{\downarrow}({\bf r})$,
which enters
\begin{equation}
\label{eq-ves}
v_{\text{es}}[n_{\text{tot}}]({\bf r})=\int d\, {\bf r'}\,
\frac{n_{\text{tot}}({\bf r'})}{|{\bf r'-r}|} \, ,
\end{equation}
is the sum
of the
spin-projected densities
\begin{equation}
\label{eq-dens}
n_{\sigma}({\bf r})=\sum_{a=1}^{N_{\sigma}} |\phi_{a\sigma}({\bf r})|^2 \, .
\end{equation}
In the HF equation
\begin{equation}
\label{eq-hf}
\hat{h}_{\text{HF}\sigma}({\bf r})  \phi_{a\sigma}^{\text{HF}}({\bf r}) \equiv
    \Bigl[ -\frac{1}{2}\bm{\nabla}^2 + v_{\text{ext}}({\bf r})+v_{\text{es}}^{\text{HF}}({\bf r})+
    \hat{v}_{\text{x}\sigma}^{\text{F}}({\bf r}) \Bigr]
    \phi_{a\sigma}^{\text{HF}}({\bf r}) =
\epsilon_{a\sigma}^{\text{HF}} \phi_{a\sigma}^{\text{HF}}({\bf r})\, ,
\end{equation}
satisfied by the orbitals
$\phi_{a\sigma}^{\text{HF}}(\bf r)$ and energies
$\epsilon_{a\sigma}^{\text{HF}}$,
the multiplicative local exchange potential
$v_{\text{x}\sigma}({\bf r})$, present in the KS equation
(\ref{eq-ks}),   is replaced with
the non-local Fock exchange integral operator
$\hat{v}_{\text{x}\sigma}^{\text{F}}({\bf r})$,
built of $\{ \phi_{a\sigma}^{\text{HF}} \}_{a=1}^{N_{\sigma}}$;
its action on a given HF orbital $\phi_{a\sigma}^{\text{HF}}$ yields \cite{johnson07}
\begin{equation}
\label{eq-fock-op-hf}
\hat{v}_{\text{x}\sigma}^{\text{F}}\left[ \{ \phi_{b\sigma}^{\text{HF}} \}\right] \! ({\bf r}) \,
\phi_{a\sigma}^{\text{HF}}({\bf r}) =
-\sum_{b=1}^{N_\sigma} \phi_{b\sigma}^{\text{HF}} ({\bf r}) \int d {\bf r'}\,
\frac{\phi_{b\sigma}^{\text{HF}}({\bf r}')\phi_{a\sigma}^{\text{HF}}({\bf r}')}{|{\bf r}'-{\bf r}|} \, .
\end{equation}
The electrostatic potential
$v_{\text{es}}^{\text{HF}}({\bf r})=v_{\text{es}}[n_{\text{tot}}^{\text{HF}}]({\bf r})$
is found
for the HF total electron density  $n_{\text{tot}}^{\text{HF}} ({\bf r})$
defined in a similar way as  $n_{\text{tot}}({\bf r})$.
The KS and HF orbitals are ordered 
according to non-descending values of the corresponding orbital energies
( $\epsilon_{a\sigma} \leq \epsilon_{a+1,\sigma}\,$ and
$\epsilon_{a\sigma}^{\text{HF}} \leq \epsilon_{a+1,\sigma}^{\text{HF}}$, $a=1,\ldots$).
Both
the KS and HF equations need to be solved selfconsistently.
Real KS and HF orbitals are used throughout this paper.

Obviously, for each HF orbital  $\phi_{a\sigma}^{\text{HF}}(\bf r)$,
the Fock exchange operator $\hat{v}_{\text{x}\sigma}^{\text{F}}({\bf r})$
present in the HF equation (\ref{eq-hf})
can be formally replaced   by $v_{\text{x}a\sigma}^{\text{HF}}({\bf r})$, Eq. (\ref{eq-vxa-hf}),
however,  this local exchange potential is orbital-dependent
due the non-locality of $\hat{v}_{\text{x}\sigma}^{\text{F}}({\bf r})$,  Eq. (\ref{eq-fock-op-hf}).
Thus, also, the resulting total HF potential
\begin{equation}
\label{eq-vsa-hf}
v_{\text{s}a\sigma}^{\text{HF}}({\bf r}) =
v_{\text{ext}}({\bf r})+v_{\text{es}}^{\text{HF}}({\bf r})+v_{\text{x}a\sigma}^{\text{HF}}({\bf r})
\end{equation}
is different for each orbital $\phi_{a\sigma}^{\text{HF}}({\bf r})$, unlike
in the  KS scheme where all electrons (of given spin $\sigma$) are subject to  the same total potential
$v_{\text{s}\sigma}({\bf r})$, which includes the common
exchange potential $v_{\text{x}\sigma}({\bf r})$.
Dependence on $\sigma$ will be suppressed hereafter (unless otherwise stated).

\subsection{Orbital and energy shifts. Exact exchange potential}
\label{sec-vxoep-os}

The exact exchange potential
$v_{\text{x}}=v_{\text{x}}^{\text{OEP}}$ satisfies the OEP equation \cite{GKKG00,KP03}
\begin{equation}
\delta n({\bf r}) \equiv  2 \sum_{a=1}^N \phi_a({\bf r})\delta\phi_a({\bf r})=0
\; ,\;\;  \forall{\bf r}\;,
\label{eq-oep}
\end{equation}
which results from the OEP minimization
and depends on $v_{\text{x}}$ through the orbital shifts (OS) $\delta \phi_a({\bf r})$.
Each OS fulfills   the equation \cite{GKKG00, KP03,CH07}

\begin{equation}
\Bigl[ \hat{h}_{\text{s}}({\bf r})-\epsilon_a \Bigr]\delta \phi_a({\bf r}) =W_a^{\perp}({\bf r})
\label{diff-eq-dphi}
\end{equation}
(where $\phi_a$, $\epsilon_a$ are the solutions of Eq. (\ref{eq-ks}))
and it is subject to the constraint $\langle \phi_a|\delta\phi_a\rangle =0$.
The equation (\ref{diff-eq-dphi}) includes
the   KS Hamiltonian $ \hat{h}_{\text{s}}$, present in Eq. (\ref{eq-ks}), and
the term (defined using the sign convention of Refs. \onlinecite{KP03, CH07})
\begin{equation}
W_a^{\perp}({\bf r})=
\left[\hat{v}_{\text{x}}^{\text{F}}({\bf r})
+D_{aa}-v_{\text{x}}({\bf r}) \right] \phi_a({\bf r}) \, .
\label{w-perp}
\end{equation}
where 
$\hat{v}_{\text{x}}^{\text{F}}=
\hat{v}_{\text{x}}^{\text{F}}\left[\{ \phi_b \}\right]$ and
\begin{equation}
\label{eq-daa}
D_{aa}= \langle \phi_a |v_{\text{x}}-\hat{v}_{\text{x}}^{\text{F}}| \phi_a\rangle \, .
\end{equation}
It should be noted that $\int d{\bf r} \, \phi_a({\bf r})W_a^{\perp}({\bf r}) =0$.

The OS $\delta\phi_a$ and the energy shift (ES) -- the constant $D_{aa}$ give, within the
perturbation theory (PT), the first-order approximations  to the
orbital and energy differences (shifts), $-(\tilde{\phi}_a^{\,\text{HF}}-\phi_a)$ and
$-(\tilde{\epsilon}_{a}^{\,\text{HF}}-\epsilon_{a})$, respectively.
Here, the
orbitals $\tilde{\phi}_a^{\,\text{HF}}$ and the corresponding
energies $\tilde{\epsilon}_a^{\,\text{HF}}$, are the solutions of
the HF-like equation which is the same as Eq. (\ref{eq-ks}) except
for $v_{\text{x}}$ replaced by $\hat{v}_{\text{x}}^{\text{F}}$ built
of the KS orbitals $\{ \phi_b \}$. The corresponding perturbation is
then equal to
$\delta\hat{h}_{\text{s}}=\hat{v}_{\text{x}}^{\text{F}}-v_{\text{x}}$
so that the first-order correction to $\epsilon_{a}$ is
$-\delta \epsilon_{a}=\langle\phi_a |\delta\hat{h}_{\text{s}}|\phi_a \rangle = - D_{aa}$ while
the  correction to $\phi_a({\bf r})$  is
\begin{eqnarray}
\label{eq-dphi}
 -\delta \phi_a ({\bf r}) & = & \sum_{j=1,\epsilon_j \neq \epsilon_a}^{\infty} c_{ja}\,\phi_j({\bf r}) \, ,\\
 \label{eq-cta}
 c_{ja} & = & \frac{D_{ja} }{\epsilon_j-\epsilon_a} \, ,\\
 \label{eq-dta}
 D_{ja} & = & -\langle\phi_j |\delta\hat{h}_{\text{s}}|\phi_a \rangle =
 \langle\phi_j |v_{\text{x}} - \hat{v}_{\text{x}}^{\text{F}}|\phi_a \rangle \, .
\end{eqnarray}
It satisfies Eq. (\ref{diff-eq-dphi}) and the constraint  $\langle \phi_a|\delta\phi_a\rangle =0$ indeed.
 Obviously, the solutions $\tilde{\phi}_a^{\,\text{HF}}$,  $\tilde{\epsilon}_a^{\,\text{HF}}$
are not identical to the selfconsistent
HF orbitals $\phi_a^{\text{HF}}$ and orbital energies $\epsilon_a^{\text{HF}}$
which are obtained from Eq. (\ref{eq-hf}).
The latter HF quantities can also be found within the PT approach
by calculating the differences $\Delta \phi_a \equiv \phi_a^{\text{HF}}-\phi_a$,
 $\Delta \epsilon_a \equiv \epsilon_a^{\text{HF}}-\epsilon_a$ in the first-order approximation.
In this case, the perturbation is given by $\Delta \hat{h}_{\text{s}} =\hat{h}_{\text{HF}} - \hat{h}_{\text{s}}$
[where $\hat{h}_{\text{HF}}$ is the HF Hamiltonian  of Eq. (\ref{eq-hf})]
and it consists of three terms,
$\Delta \hat{h}_{\text{s}} =
\delta \hat{h}_{\text{s}}  +\Delta v_{\text{es}} + \Delta \hat{v}_{\text{x}}^{\text{F}}\;$.
The terms
$\Delta v_{\text{es}}= v_{\text{es}}[n_{\text{tot}}^{\text{HF}}] - v_{\text{es}}[n_{\text{tot}}]
=v_{\text{es}}[n_{\text{tot}}^{\text{HF}} - n_{\text{tot}}]$
(cf. Eq. (\ref{eq-ves}))
 and
$\Delta \hat{v}_{\text{x}}^{\text{F}}= \hat{v}_{\text{x}}^{\text{F}}[\{\phi_a^{\text{HF}}\}] -
                                                     \hat{v}_{\text{x}}^{\text{F}}[\{\phi_a\}] $
depend on $\Delta \phi_a$ (of both spins for $\Delta v_{\text{es}}$), linearly in the leading order,
so that they have to be calculated selfconsistently even in the PT approach.
But, if we substitute $(-\delta \phi_a)$ for $\Delta \phi_a$ 
the difference
$n_{\text{tot}}^{\text{HF}} - n_{\text{tot}}$ becomes  $\delta n_{\uparrow}+\delta n_{\downarrow}$
so that it vanishes due to the OEP equation (\ref{eq-oep}).
Then, we find $\Delta v_{\text{es}}=0$  and the perturbation $\Delta \hat{h}_{\text{s}}$ becomes
$\delta \hat{h}_{\text{s}}  + \Delta \hat{v}_{\text{x}}^{\text{F}}\left[ \{\phi_a\}, \{\delta\phi_a\}\right]$.
It can be further reduced to $\delta \hat{h}_{\text{s}}$  if the OS $\delta\phi_a$ are sufficiently small.
This argument, although not strict, leads to the conclusion that the differences
$\Delta \phi_a $ and  $\Delta \epsilon_a$  are well represented by
the orbital and energy shifts,
$-\delta\phi_a$ and $-\delta\epsilon_a=-D_{aa}$, respectively, which are obtained with
the perturbation $\delta \hat{h}_{\text{s}}$.
This conclusion is confirmed by
the relations $\| \Delta \phi_a - (-\delta\phi_a) \| < 0.13 \| \Delta \phi_a \|$
(where $\| \phi\|^2=\int d {\bf r} \,|\phi({\bf r})|^2$) and
$| \Delta \epsilon_a - (-\delta\epsilon_a) | < 0.003 | \Delta \epsilon_a |$
\cite{note-diff-en},  established numerically for the Be and Ar atoms (see Tables \ref{tab-dphi}
and \ref{tab-dnl});  
the above inequalities are obtained for $\phi_a$, $\epsilon_a$,
$\delta\phi_a$ calculated as in 
Ref. \cite{CH07}, and
$\phi_a^{\text{HF}}$ (expanded in the Slater-type-orbital basis), $\epsilon_a^{\text{HF}}$
taken from Ref. \cite{bunge93}.
The representations of $\phi_a^{\text{HF}}-\phi_a$ by $-\delta\phi_a$
and $\epsilon_a^{\text{HF}}-\epsilon_a$ by $-\delta\epsilon_a$
will be used  in  further discussion.
They can also be applied to construct a nearly accurate approximation of the exact exchange
potential; the new method will be reported elsewhere soon \cite{CH09}.

The part of
$W_a({\bf r}) \equiv \delta \hat{h}_{\text{s}}({\bf r})\phi_a({\bf r})$
parallel to the orbital $\phi_a$ is
\begin{equation}
\label{eq-w-par}
W_a^{||}({\bf r})=-D_{aa}\phi_a({\bf r})
\end{equation}
and it sets the ES $\delta \epsilon_a=D_{aa}$.
The part
\begin{equation}
\label{eq-w-perp2}
W_a^{\perp}({\bf r})=W_a({\bf r})-W_a^{||}({\bf r}) \, ,
\end{equation}
perpendicular to $\phi_a$,
sets  the OS $\delta \phi_a({\bf r})$, Eqs. (\ref{diff-eq-dphi}), (\ref{w-perp}).
Thus, the KS and HF orbitals, $\phi_a({\bf r})$, $\phi_a^{\text{HF}}({\bf r})$,
can be close to each other, even if the orbital energies
$\epsilon_{a}$, $\epsilon_{a}^{\text{HF}}$,
differ significantly, provided  the term $W_a^{\perp}({\bf r})$ is sufficiently small.
Note that the orbitals remain unchanged when a (possibly orbital-dependent) constant
is added to the Hamiltonian in the KS or HF  equations.

When the equation (\ref{diff-eq-dphi}) (after multiplying it  by $\phi_a({\bf r})$ and subsequent summing
over $a=1,\ldots,N$) is combined with the OEP condition (\ref{eq-oep}),
the following expression  \cite{GKKG00, KP03,CH07} for
the exact exchange potential is obtained
\begin{equation}
\label{eq-vx-oep}
v_{\text{x}}^{\text{OEP}}({\bf r}) =\breve{v}_{\text{x}}^{\text{KLI}}({\bf r})  +v_{\text{x}}^{\text{OS}}({\bf r})
\end{equation}
It contains
the KLI-like potential \cite{KLI92}
\begin{equation}
\label{eq-vx-kli-oep}
\breve{v}_{\text{x}}^{\text{KLI}} \left[ \{\phi_a\}, \{D_{aa}\}\right] ({\bf r})=
v_{\text{x}}^{\text{Sl}}({\bf r}) + v_{\text{x}}^{\text{ES}}({\bf r})
\end{equation}
which
consists of the Slater potential
\begin{equation}
\label{eq-vx-slater}
v_{\text{x}}^{\text{Sl}}({\bf r})=
\frac {1}{n({\bf r})} \sum_{a=1}^{N} \phi_a({\bf r}) \hat{v}_{\text{x}}^{\text{F}}({\bf r})\phi_a({\bf r})
\end{equation}
and the ES term, linear in $D_{aa}$,
\begin{equation}
\label{eq-vx-es}
v_{\text{x}}^{\text{ES}}({\bf r}) =
\frac {1}{n({\bf r})} \sum_{a=1}^{N} D_{aa} \phi_a^2({\bf r})
\end{equation}
where $n({\bf r})= \sum_{a=1}^{N} \phi_a^2({\bf r})$ ;
these terms are defined with  the OEP orbitals $\phi_a({\bf r})$ and  constants $D_{aa}$.
The OS term present in Eq. (\ref{eq-vx-oep}), linear in $\delta\phi_a({\bf r})$,  is
\begin{equation}
\label{vx-os}
v_{\text{x}}^{\text{OS}}({\bf r}) = \frac{1}{n({\bf r})}
\sum_{a=1}^{N} \bigl[ 2 \epsilon_a \phi_a({\bf r}) -
\bigl(\bm{\nabla} \phi_a({\bf r})\bigr)\!\cdot\!\bm{\nabla}
\bigr]\delta\phi_a({\bf r}) \, .
\end{equation}
Since any physical potential is defined up to an arbitrary constant,
it is usually chosen that the constant $D_{NN}=0$ for the HOMO \cite{KP03};
then the potential $v_{\text{x}}^{\text{OEP}}({\bf r})$ goes to 0 as $-1/r$ for $r=|{\bf r}| \rightarrow \infty$
(except for the directions that lie within symmetry planes in some molecules: in this special case
the $(-1/r+{\text{const}})$ dependence at large $r$ is found; cf.  Ref. \onlinecite{KP03,DSG02}).

However, the use of  Eq. (\ref{eq-vx-oep}) for calculation of $v_{\text{x}}^{\text{OEP}}({\bf r})$
still requires solving the equations (\ref{eq-oep},\ref{diff-eq-dphi}) for  $\delta\phi_a({\bf r})$ as well as
determining   the selfconsistent values of the constants $D_{aa}$
which depend on $v_{\text{x}}=v_{\text{x}}^{\text{OEP}}({\bf r})$ through Eq. (\ref{eq-daa}).
This solution is obtained in an iterative way in Ref. \onlinecite{KP03},
while a non-iterative algorithm, where both sets $\{\delta\phi_a\}$ and $\{D_{aa}\}$
are found simultaneously, is presented in Ref. \onlinecite{CH07}.
Let us note that the equations
(\ref{eq-oep}-\ref{eq-daa}),
(\ref{eq-vx-oep}-\ref{vx-os})
can be used to determine the exact exchange potential $v_{\text{x}}^{\text{OEP}}({\bf r})$
not only in the exchange-only OEP approach, but also when the orbitals
$\phi_a({\bf r})$ are the solutions of the KS equation  with the potential
$v_{\text{s}}({\bf r})$ that, besides $v_{\text{x}}({\bf r})$,
includes a correlation term $v_{\text{c}}({\bf r})$.

\subsection{High-quality KLI and LHF (CEDA) approximations}
\label{sec-vx-kli-lhf}

Since the OS  $\delta\phi_a({\bf r})$ are usually small,
the term  $v_{\text{x}}^{\text{OS}}({\bf r})$, Eq. (\ref{vx-os}),
is a minor correction to $v_{\text{x}}^{\text{KLI}}({\bf r})$ in Eq. (\ref{eq-vx-oep}).
Therefore, when we  neglect  $v_{\text{x}}^{\text{OS}}({\bf r})$  completely,
 the exact exchange potential $v_{\text{x}}^{\text{OEP}}({\bf r})$ is represented with high quality
by the KLI-like term $\breve{v}_{\text{x}}^{\text{KLI}}({\bf r})$, Eq. (\ref{eq-vx-kli-oep}).
The original KLI  approximation  \cite{KLI92}
\begin{equation}
\label{eq-vx-kli-orig}
v_{\text{x}}^{\text{KLI}}\left[ \{\phi_a\} \right]({\bf r}) =
\breve{v}_{\text{x}}^{\text{KLI}} \left[ \{\phi_a\}, \{D_{aa}^{\text{KLI}}\}\right]({\bf r})
\end{equation}
is obtained (here for the KS-OEP orbitals $\phi_a$)
when the constants
\begin{equation}
\label{daa-kli}
D_{aa}^{\text{KLI}}=\langle \phi_a |v_{\text{x}}^{\text{KLI}}-\hat{v}_{\text{x}}^{\text{F}}| \phi_a\rangle
\end{equation}
are found  selfconsistently, analogously as $D_{aa}$ in Eq. (\ref{eq-daa})
for $v_{\text{x}}=v_{\text{x}}^{\text{OEP}}$.
Since, the equation (\ref{daa-kli}) remains satisfied when an arbitrary constant, but  the same for all $a$,
is added to each $D_{aa}^{\text{KLI}}$, one usually sets
$D_{NN}^{\text{KLI}}=0$ which makes the potential $v_{\text{x}}^{\text{KLI}}({\bf r})$ decay like
$-1/r$ for large $r$.

The sum over $j$ in Eq. (\ref{eq-dphi}) can be split  into two terms,
\begin{eqnarray}
\label{eq-dphi-occ}
\delta \phi_a^{\text{occ}}  & = &  -\sum_{b=1,b \neq a}^{N} c_{ba}\,\phi_b \, , \\
\label{eq-dphi-vir}
\delta \phi_a^{\text{vir}}   & = &  -\sum_{t=N+1}^{\infty} c_{ta}\,\phi_t \, ,
\end{eqnarray}
which are the projections of the OS $\delta\phi_a$ onto the subspaces of occupied (occ)
and virtual (vir) orbitals, respectively.
Thus, the OS term $v_{\text{x}}^{\text{OS}}({\bf r})$, Eq. (\ref{vx-os}),  can be rewritten as follows
\begin{equation}
\label{eq-vx-os2}
v_{\text{x}}^{\text{OS}} ({\bf r}) =
\frac{1}{n({\bf r}) } \sum_{a=1}^{N-1}\sum_{b=a+1}^{N} 2 D_{ab} \phi_a({\bf r})\phi_b({\bf r})+
v_{\text{x}}^{\text{OS,vir}}({\bf r})
\end{equation}
after the definition (\ref{eq-cta}) of $c_{ta}$ and relation $D_{ba}=D_{ab}$
[cf. Eq. (\ref{eq-dta}), $v_{\text{x}}^{\text{F}}$  is Hermitian and real] are used;
the term
\begin{equation}
\label{eq-vx-os-vir}
v_{\text{x}}^{\text{OS,vir}}({\bf r})=
v_{\text{x}}^{\text{OS}}\left[ \{\phi_a\}, \{\epsilon_a\},\{\delta\phi_a^{\text{vir}} \}\right]({\bf r})
\end{equation}
is found by substituting $\delta\phi_a^{\text{vir}}$ for $\delta\phi_a$ in Eq. (\ref{vx-os}).
When the OS $\delta\phi_a$ are small, the corresponding projected parts $\delta\phi_a^{\text{vir}}$
are even smaller since the general relation
$\| \delta\phi_a^{\text{occ}}\|^2 + \| \delta\phi_a^{\text{vir}}\|^2 = \| \delta\phi_a\|^2$ holds.
Then,  another high-quality representation
of   $v_{\text{x}}^{\text{OEP}}$
\begin{eqnarray}
\label{eq-vx-lhf}
 \breve{v}_{\text{x}}^{\text{LHF}}\left[ \{\phi_a\}, \{D_{ab}\} \right] ({\bf r}) =
v_{\text{x}}^{\text{Sl}} ({\bf r}) +\frac{1}{n({\bf r}) } \sum_{a,b=1}^{N} D_{ab} \phi_a({\bf r})\phi_b ({\bf r})
\end{eqnarray}
is obtained by setting $\delta\phi_a^{\text{vir}}({\bf r})=0$  in
the OS term $v_{\text{x}}^{\text{OS}}({\bf r})$, Eqs. (\ref{eq-vx-os2}, \ref{eq-vx-os-vir}).
This representation
yields the well-known LHF (CEDA) approximation \cite{GB01,DSG01}
\begin{equation}
\label{eq-vx-lhf-orig}
 v_{\text{x}}^{\text{LHF}}\left[ \{\phi_a\} \right]
= \breve{v}_{\text{x}}^{\text{LHF}}\left[ \{\phi_a\}, \{D_{ab}^{\text{LHF}}\} \right]
\end{equation}
(here defined for the set $\{\phi_a\}$ of the KS-OEP orbitals) when
the constants
$D_{ab}^{\text{LHF}}=\langle \phi_a |v_{\text{x}}^{\text{KLI}}-\hat{v}_{\text{x}}^{\text{F}}| \phi_b\rangle \, ,$
defined analogously as in  Eq. (\ref{eq-dta}), are found selfconsistently for $(ab)\neq (NN)$;
we also set $D_{NN}^{\text{LHF}}=0$, as in the KLI case.
Let us note that
the condition $\delta\phi_{a\sigma}^{\text{vir}}=0$ is equivalent to the relation
 $\tilde{\phi}_{a\sigma} ^{\text{HF}}=\phi_{a\sigma} +\sum_{b \neq a}^{\text{occ}}c_{ba} \phi_{b\sigma}$
(valid in the first-order approximation) which, when satisfied for both spins $\sigma$,
implies that the HF determinant built of $\{\tilde{\phi}_{a\sigma}^{\text{HF}}\}$ is
identical to the KS determinant  built of $\{ \phi_{a\sigma} \}$.
This (approximate) identity has been assumed in Ref. \onlinecite{DSG01} to derive the LHF
approximation.
Obviously, both the KLI and LHF approximate exchange potentials
can be defined for any set of (orthogonal, bound) orbitals $\{\phi_a\}_{a=1}^N$.
In particular, it can be done for the orbitals
that are selfconsistent solutions of the KS equation (\ref{eq-ks}) where the potential
$v_{\text{x}}$  is set to $v_{\text{x}}^{\text{KLI}}\left[ \{\phi_a\} \right]$ or
$v_{\text{x}}^{\text{LHF}}\left[ \{\phi_a\} \right]$.

The high quality of the KLI and LHF approximate potentials,
when derived as presented above, clearly results from
the proximity of the HF and KS-OEP occupied orbitals
which is characterized  by the small OS $\delta \phi_a$.
However, the OS terms
$v_{\text{x}}^{\text{OS}}$ and $v_{\text{x}}^{\text{OS,vir}}$
which are neglected in the KLI and LHF (CEDA) approximations, respectively, are expressed
through  {\em all} OS $\delta \phi_a$ (or their projected parts $\delta\phi_a^{\text{vir}}$).
As a result,  some information associated with
the small magnitudes of the {\em individual} OS $\delta \phi_a$ may be lost
in the resulting potentials $v_{\text{x}}^{\text{KLI}}$ and
$v_{\text{x}}^{\text{LHF}}$.
In particular,   the Slater term, Eq. (\ref{eq-vx-slater}), present in these  potentials,
can be viewed the weighted average
\begin{equation}
\label{eq-vx-slater-av-vxa}
v_{\text{x}}^{\text{Sl}}({\bf r})=
\sum_{a=1}^{N} v_{\text{x}a}({\bf r}) \frac {\phi_a^2({\bf r})}{n({\bf r})}
\end{equation}
of the KS orbital exchange potentials
\begin{equation}
\label{eq-vxa-ks}
v_{\text{x}a}({\bf r})=\frac{\hat{v}_{\text{x}}^{\text{F}}({\bf r})\phi_a({\bf r})}{\phi_a({\bf r})} \, ,
\end{equation}
so that it {\em cannot  fully}  reflect the properties of
the individual $v_{\text{x}a}({\bf r})$.
In the following discussion (Sec. \ref{sec-results}) for closed-$l$-subshell atoms,
new properties of $v_{\text{x}a}({\bf r})$ are exposed {\em only} when 
the proximity of the HF and KS-OEP orbitals is considered  {\em separately} for each orbital.

\subsection{Closed-$l$-subshell atoms: Fock exchange operator, orbital exchange potentials }
\label{sec-theory-closed-shell-atoms}

For a closed-$l$-subshell atom,
the non-local (integral) Fock exchange operator,
acting on an atomic orbital
$\phi_a({\bf r})=r^{-1}\chi_{nl}(r) Y_{lm}(\theta,\varphi)$ ($a\equiv nlm$),
yields
\begin{equation}
\label{eq-vxfock-ylm}
\hat{v}_{\text{x}}^{\text{F}}({\bf r})\phi_a({\bf r}) =
r^{-1} F_{\text{x};nl}(r) \, Y_{lm}(\theta,\phi)
\end{equation}
where  $Y_{lm}(\theta,\varphi)$ is the spherical harmonic,
Hereafter, the orbitals are labeled with   the principal,  orbital,
and magnetic quantum numbers, $n$, $l$, $m$;
the symbols $n_{\text{occ}}$ and $l_{\text{max}}^{(n)}$ will denote, respectively,
the largest  number $n$ and the maximum value of $l$ for given $n$,
within the set $\{\text{occ}\}$  
of  the occupied orbitals $\{ \phi_{nlm} \}$
(hereafter, we refer to this set with the general label "occ").
It will be convenient to have a notation for the HOMO label:
$H\equiv (n\,l_{\text{max}}^{(n)})$ at $n=n_{\text{occ}}$; note that
the HOMO belongs to the outmost occupied shell for  the closed-$l$-shell atoms.
The factor
\begin{subequations}
\label{eq-fxnl}
\begin{equation}
\label{eq-fxnl-gen}
F_{\text{x};nl}(r) =
 \sum_{n'l'}^{\text{occ}} \sum_{l''=|l-l'|}^{l+l'}
g(l,l',l'') \chi_{n'l'}(r) \, v_{l''}(n'l',nl;r) \, ,
\end{equation}
is defined  \cite{johnson07}
(here with the occupied KS radial orbitals $\chi_{n'l'}(r)\,$) through the functions 
\begin{equation}
\label{eq-fxnl-vl}
 v_{l''}(n'l',nl;r) =  - \int_{0}^{\infty} dr' \frac{(r_{<})^{l''}}{(r_{>})^{l''+1}} \chi_{n'l'}(r')\chi_{nl}(r')
\end{equation}
\end{subequations}
where we denote
$g(l,l',l'')=(2l'+1) 
\left(
\begin{array}{c c c}
 l & l' & l''  \\
0 & 0 & 0 \\
\end{array} \right)
 $
(a special case of the $3j$ Wigner symbol), $r_{<}=\min(r,r')$, $r_{>}=\max(r,r')$.
In particular, the following non-zero coefficients
$g(0,0,0)=1$, $g(0,1,1)=1$, $g(1,1,0)=1$, $g(1,0,1)=1/3$, $g(1,1,2)=2/5$
are needed to find the quantities $F_{\text{x};nl}(r)$
for atoms with $s$ and $p$ orbitals (like Be, Ar); note that the step in
the summation over $l''$  in  Eq. (\ref{eq-fxnl-gen})  is 2.
Thus, the orbital exchange potential, Eq. (\ref{eq-vxa-ks}),
\begin{equation}
\label{eq-vxnl}
v_{\text{x}a}({\bf r})=v_{\text{x};nl}(r)=
F_{\text{x};nl}(r)/\chi_{nl}(r) \, ,
\end{equation}
is obtained;
the corresponding HF quantities, denoted as
$v_{l''}^{\text{HF}}(n'l',nl;r)$,   $F_{\text{x};nl}^{\text{HF}}(r)$,
$v_{\text{x}a}^{\text{HF}}({\bf r})=v_{\text{x};nl}^{\text{HF}}(r)$, can
be determined with the HF atomic radial orbitals $\chi_{nl}^{\text{HF}}(r)$.
The OS
\begin{equation}
\label{eq-dphi-dchi}
\delta\phi_a({\bf r})=r^{-1}\delta\chi_{nl}(r) Y_{lm}(\theta,\varphi)
\end{equation}
depends on  the term
\begin{equation}\label{eq-waperp}
W_a^{\perp}({\bf r})=r^{-1}W_{nl}^{\perp\text{;rad}}(r) Y_{lm}(\theta,\varphi)
\end{equation}
through its radial part
\begin{equation}
\label{wrad-perp}
W_{nl}^{\perp ;\text{rad}}(r)=F_{\text{x};nl}(r)+\left[D_{nl;nl}-v_{\text{x}}(r)\right]\chi_{nl}(r)
\end{equation}
entering the equation
\begin{equation}
\label{diff-eq-dchi}
\left [ -\frac{1}{2} \frac{d^2}{dr^2}+\frac{l(l+1)}{2r^2} +
v_{\text{s}}(r) -\epsilon_{nl}\right ]\delta \chi_{nl}(r)  =
W_{nl}^{\perp ;\text{rad}}(r) \chi_{nl}(r)
\end{equation}
for $\delta\chi_{nl}(r)$ derived from Eq. (\ref{diff-eq-dphi});
here $\epsilon_{nl}$ is the energy of the KS orbital $\phi_a=\phi_{nlm}$.
The KS potential $v_{\text{s}}(r)$, Eq. (\ref{eq-vs-ks}) contains
the term $v_{\text{ext}}(r)=-Z/r$ where  $Z$ is the atomic number,
equal to $N$ for neutral atoms.

\section{NUMERICAL RESULTS AND DISCUSSION}
\label{sec-results}

\subsection{Proximity of KS and HF orbitals}
\label{sec-results-proximity}

%
The proximity of individual HF and KS orbitals can be quantified
with the norms $\|\delta \phi_a\|$ which are found to be indeed very small,
in comparison with
$\|\phi_a\|=\|\phi_a^{\text{HF}}\|=1$.
Calculating the  OS $\delta\phi_a$  with the method of Ref. \onlinecite{CH07},
we  obtain 
$\|\delta \phi_a\|<0.007$ for
each occupied orbital in the Be and Ar atoms; see Table \ref{tab-dphi}.
The partition
\begin{equation}
\label{eq-os-exp}
\|\delta \phi_{nlm}\|^2=\sum_{n' \neq n}^{\infty} c_{n'l;nl}^2 \, ,
\end{equation}
plotted for Ar in Fig. \ref{fig-dchi-dist-Ar},
shows that, among the  KS {\em bound} orbitals $\phi_{n'lm}$,
the dominating contributions $c_{n'l;nl}^2$ to the $nlm$ OS
come from
the  $n'lm$ orbitals with $n'=n-1$ and/or $n+1$, i.e., from
the neighboring electronic shells;
e.g., for  $\delta\phi_{3s}$ in the Ar atom, the largest terms
$c_{n'l;nl}^2$ are found for the $n'l=2s$ (occupied) and $n'l=4s$ (unoccupied) orbitals.
But,  there remains a large  part
of $\|\delta \phi_{nlm}\|^2$
which cannot be attributed to higher unoccupied bound
states $\phi_{n'lm}$ since
the corresponding $c_{n'l;nl}^2$ terms vanish rapidly with increasing $n'$;
see  Fig. \ref{fig-dchi-dist-Ar}.
This unaccounted part comes from {\em continuum} KS states
($\epsilon_{n'l}>0$).
Let us also note that, for each OS $\delta \phi_{nlm}$ analyzed
in  Fig. \ref{fig-dchi-dist-Ar},
its projection $\delta \phi_{nlm}^{\text{occ}}$, Eq. (\ref{eq-dphi-occ}),
onto the occupied-state subspace has the squared norm smaller than
$\frac{1}{2}\|\delta \phi_{nlm}\|^2$ which means that the relation
$\|\delta \phi_{nlm}^{\text{occ}}\| < \|\delta \phi_{nlm}^{\text{vir}}\|$
holds for the Ar atom.

The above  results also confirm that the assumptions
$\delta\phi_{a}=0$ and
$\delta \phi_a ^{\text{vir}}=0$, which can be
used to derive the KLI and LHF (CEDA) approximations,
respectively (cf. Sec. \ref{sec-vx-kli-lhf}), are very accurate but not exact.

\subsection{Exact exchange potential vs orbital exchange potentials}
\label{sec-results-vxoep-vxa}

The norms $\|\delta \phi_{nlm}\|$ have such low values
because the terms $W_{nl}^{\perp;\text{rad}}(r)$
are sufficiently small {\em for all} $r$
(the scale of this smallness will be discussed later on). 
This, combined with the relation
\begin{equation}
v_{\text{x;}nl}(r)+D_{nl;nl}=
v_{\text{x}}(r)+
\frac{W_{nl}^{\perp\text{;rad}}(r)}{\chi_{nl}(r)}\; ,
\label{vxa-tilde}
\end{equation}
found with Eqs. (\ref{wrad-perp}) and (\ref{eq-vxnl}), implies that each
{\em shifted}  {\em orbital} exchange potential (calculated from the KS-OEP orbitals)
\begin{equation}
\label{vxa-tilde-def}
\tilde{v}_{\text{x;}nl}(r)\equiv v_{\text{x;}nl}(r)+D_{nl;nl}
\end{equation}
is very close to the exact exchange potential
$v_{\text{x}}(r)=v_{\text{x}}^{\text{OEP}}(r)$ within the
$r$-interval $(r_{n-1,n},r_{n,n+1})\equiv S_n$ where the
denominators in the right-hand side of Eq. (\ref{vxa-tilde}), i.e., the
orbitals $\chi_{nl}(r)$ from the $n$-th atomic shell
($K,L,M,\ldots$), have largest magnitudes. The shell border points
$r_{n,n+1}$ for $n=0,1,\ldots,n_{\text{occ}}-1$ (the
respective HF points $r_{n,n+1}^{\text{HF}}$, defined precisely
below, can be used) are near the positions $r_{n}^{\text{min}}$ where the radial
electron density $\rho(r)$ has local minima.
In large parts 
of the shells $S_{n'}$,  $n'<n$,
where the orbital $\chi_{nl}(r)$ entering the denominator in Eq. (\ref{vxa-tilde})
has sizeable magnitude (though at least a few times smaller  than
in the shell $S_n$) the potentials $\tilde{v}_{\text{x;}nl}(r)$ are also
close to  $v_{\text{x}}^{\text{OEP}}(r)$ (but not so tightly as for $r\in S_n$).
The proximity of the potentials is evident in Figs.
\ref{fig-Be_vx_vxa}, \ref{fig-Ar_vx_vxa}, \ref{fig-Zn_vx_vxa} for the Be, Ar, and Zn atoms, respectively;
it also holds for other closed-$l$-subshell atoms.
It is disturbed  in  the vicinity of the nodes of $\chi_{nl}(r)$,
where the potential $\tilde{v}_{\text{x;}nl}(r)$ diverges while
the  term
$W_{nl}^{\perp\text{;rad}}(r)$
is finite and small.
The potential  $\tilde{v}_{\text{x;}nl}(r)$ also differs significantly from
 $v_{\text{x}}^{\text{OEP}}(r)$ within the occupied shells
$S_{n''}$, $n''>n$, where both the functions $\chi_{nl}(r)$, $W_{nl}^{\perp\text{;rad}}(r)$
decay exponentially.

In the asymptotic region $S_{\infty}$ (spanning outside the  occupied shells, i.e.,
for $r>r_{n,n+1}$, $n=n_{\text{occ}}$) the exact exchange potential $v_{\text{x}}^{\text{OEP}}(r)$
lies very close {\em only} to the  HOMO exchange potential $\tilde{v}_{\text{x;}H}(r)=v_{\text{x;}H}(r)$
(where $D_{H,H}=0$) which has the correct  $-1/r$ dependence for large $r$ resulting from
Eqs. (\ref{eq-fxnl}), (\ref{eq-vxnl}); see Fig. \ref{fig-Be_Ar-asymp}.
Indeed, the potential $\tilde{v}_{\text{x;}nl}(r)$  for  $nl \neq H$
includes, besides  the self-interaction term  $v_{0}(nl,nl;r)$, equal to $-1/r$
for large $r$, also, at least  one non-zero term proportional to
$\chi_{H}(r)v_{l''}(H,nl;r)/\chi_{nl}(r) $ with $l''\neq 0$;
cf. Eqs. (\ref{eq-fxnl}), (\ref{eq-vxnl}).
The latter term diverges for  $r\rightarrow \infty$ since
the factor $v_{l''}(H,nl;r)$ tends to a constant
while each KS radial orbital $\chi_{nl}(r)$ decays like $r^{1/\beta_{nl}}e^{-\beta_{nl}r}$
where $\beta_{nl}=\sqrt{-2\epsilon_{nl}}$ (cf. Ref. \onlinecite{GKKG00});
this is true also for $nl=H$.
The Be atom, with the $1s$ and $2s$ orbitals only, is the only exception here since,
in this case, both potentials
$v_{\text{x;}1s}(r)$ and  $v_{\text{x;}2s}(r)$  decay as $-1/r$ for large $r$.
Indeed, with
Eqs. (\ref{eq-fxnl}), (\ref{eq-vxnl}) we find  the following expressions
\begin{subequations}
\begin{equation}
\label{eq-v1s-be}
v_{\text{x;}1s}(r)=v_0(1s,1s;r)+\frac{\chi_{2s}(r)}{\chi_{1s}(r)}v_0(2s,1s;r)  \, ,
\end{equation}
\begin{equation}
\label{eq-v2s-be}
v_{\text{x;}2s}(r)=v_0(2s,2s;r)+\frac{\chi_{1s}(r)}{\chi_{2s}(r)}v_0(1s,2s;r) \, ,
\end{equation}
\end{subequations}
valid for the Be atom.
Due the orthogonality of the $1s$ and $2s$ orbitals, the function 
$v_0(1s,2s;r)=v_0(2s,1s;r)$, Eq. (\ref{eq-fxnl-vl}),  is equal to
$\int_r^{\infty} dr' (1/r-1/r') \chi_{1s}(r')\chi_{2s}(r')$ so that it decays exponentially
like $\chi_{1s}(r)\chi_{2s}(r)$ for large $r$.
Thus, the second terms in the expressions
(\ref{eq-v1s-be}), (\ref{eq-v2s-be}) for $v_{\text{x;}1s}(r)$ and $v_{\text{x;}2s}(r)$
also decay exponentially, as $\chi_{2s}^2(r)$ and $\chi_{1s}^2(r)$, respectively.
As a result, the self-interaction energies,  $v_0(1s,1s;r)$ and $v_0(2s,2s;r)$,
which both depend like $-1/r$ for large $r$,
dominate in the respective potentials $v_{\text{x;}1s}(r)$ and  $v_{\text{x;}2s}(r)$ in the
asymptotic region $S_{\infty}$.

As it is seen in Fig. \ref{fig-Be_vx_vxa}(b) for the Be atom,  the quantities
$\tilde{v}_{\text{x;}nl}(r)\chi_{nl}(r)= F_{\text{x;}nl}(r)+D_{nl;nl}\chi_{nl}(r)$
and $v_{\text{x}}(r)\chi_{nl}(r)$,
whose difference yields $W_{nl}^{\perp\text{;rad}}(r)$, Eq. (\ref{wrad-perp}),  lie close to each other
for all $r$.
However,
it is not straightforward to define a direct 
scale  that could serve to estimate
how small the potential difference
$\tilde{v}_{\text{x;}nl}(r) - v_{\text{x}}(r)$, or rather, the term
$W_{nl}^{\perp\text{;rad}}(r)$ should be to make the OS $\delta \phi_{nlm}$ small.
Indeed, it is the ratio of the overlap integrals
$D_{n'l;nl}=-\int_0^{\infty} dr' \chi_{n'l}(r')W_{nl}^{\perp\text{;rad}}(r')$
and  the orbital energy differences $\epsilon_{n'l}-\epsilon_{nl}$,
that,  in fact,  determine the expansion coefficients $c_{n'l;nl}=D_{n'l;nl}/(\epsilon_{n'l}-\epsilon_{nl})$,
and, consequently, the magnitude of the OS $\delta \phi_{nlm}$;
cf. Eqs. (\ref{eq-dphi}-\ref{eq-dta}), (\ref{eq-os-exp}).
Since the difference $\epsilon_{n',l}-\epsilon_{nl}$ (with given $l$ and $n'\neq n$) has the smallest magnitude
for $n'=n+1$, we could find an upper bound for the OS norm,
\begin{equation}
\label{eq-dphi-upper-bound}
\|\delta \phi_{nlm} \| \leq     \frac{\sum_{n'\neq n} |D_{n'l;nl}|^2}{| \epsilon_{n+1,l}-\epsilon_{nl} |} =
\frac{\|W_{nl}^{\perp\text{;rad}}\|}{| \epsilon_{n+1,l}-\epsilon_{nl} |} \, ;
\end{equation}
which is expressed, as it would be desired, in terms of the whole norm of $W_{nl}^{\perp\text{;rad}}(r)$.
However, this bound gives  values that largely exceed $\|\delta \phi_{nlm} \|$ for the considered atoms;
see Table \ref{tab-dphi}.
Thus, it seems that, ultimately,
the only fully adequate measure (in the present context) of the smallness of  $W_{nl}^{\perp\text{;rad}}(r)$
is the smallness of the norms $\|\delta \phi_{nlm} \|$
that are generated by $W_{nl}^{\perp\text{;rad}}(r)$.

\subsection{Properties of Hartree-Fock orbital exchange potentials}
\label{sec-prop-HF-vxnl}

Since the exchange-only  KS orbitals
$\phi_a({\bf r}) = \phi_a^{\text{OEP}}({\bf r})$ found with the
exact exchange potential $v_{\text{x}}^{\text{OEP}}(r)$
are very close to   $\phi_a^{\text{HF}}({\bf r})$, the terms $F_{\text{x;}nl}(r)$, $v_{\text{x;}nl}(r)$,
and $D_{nl,nl}[v_{\text{x}}]$
obtained with $\{\phi_a^{\text{OEP}}\}$
are virtually indistinguishable from
the respective quantities $F_{\text{x;}nl}^{\text{HF}}(r)$, $v_{\text{x;}nl}^{\text{HF}}(r)$,
$D_{nl,nl}^{\text{HF}}[v_{\text{x}}]$
calculated with
the HF orbitals $\{\phi_a^{\text{HF}}\}$
(it is true for {\em any} $v_{\text{x}}$ used as the argument of $D_{nl,nl}$ and $D_{nl,nl}^{\text{HF}}$).
Thus,
the combinations of these terms
\begin{eqnarray}
\label{eq-wnl}
W_{nl}^{\perp;\text{rad}}[v_{\text{x}},\{\phi_a^{\text{HF}}\}] (r) =  
F_{\text{x;}nl}^{\text{HF}}(r) + D_{nl,nl}^{\text{HF}}[v_{\text{x}}]\chi_{nl}^{\text{HF}}(r) -
v_{\text{x}}(r) \chi_{nl}^{\text{HF}}(r)
\end{eqnarray}
 are very close to
$W_{nl}^{\perp;\text{rad}}[v_{\text{x}},\{\phi_a^{\text{OEP}}\}](r)$.
As a result, they are small for
$v_{\text{x}}=v_{\text{x}}^{\text{OEP}}$
(since the  quantities
$W_{nl}^{\perp;\text{rad}}[v_{\text{x}}^{\text{OEP}},\{\phi_a^{\text{OEP}}\}]$
have been found to be small),
and, also, by continuity, for any approximate potential
$v_{\text{x}}$ close to $v_{\text{x}}^{\text{OEP}}$.
Therefore, basing on the numerically established   proximity
of the KS-OEP and HF occupied orbitals $\phi_{nlm}({\bf r})$ in closed-$l$-shell atoms,
we conclude that there exists a non-empty class ${\cal V}_0$
of approximate exchange potentials  $v_{\text{x}}$ that yield small terms
$W_{nl}^{\perp;\text{rad}}[v_{\text{x}},\{\phi_a^{\text{HF}}\}]$.
In addition, we can assume that these potentials
have correct, $-1/r$,  dependence at large $r$ and
lead to $D_{H,H}^{\text{HF}} \left[ v_{\text{x}}\right] \approx 0$
(since these two conditions are fulfilled by $v_{\text{x}}^{\text{OEP}}$).

The class ${\cal V}_0$  is constituted, in fact,
by all potentials $v_{\text{x}}$ (with correct asymptotics) for each of which
it is possible to find constants $C_{nl}$
that make terms
\begin{equation}
\label{eq-unl}
U_{nl}(r) \equiv  F_{\text{x;}nl}^{\text{HF}}(r)+C_{nl}\chi_{nl}^{\text{HF}}(r)
                       - v_{\text{x}}(r)\chi_{nl}^{\text{HF}}(r)
\end{equation}
small for {\em all} $r$ and every occupied orbital $\chi_{nl}^{\text{HF}}(r)$;
additionally,  we set $ C_{H}=0$.
Indeed, this definition (\ref{eq-unl}) allows us to write  
(cf. Eq. (\ref{eq-daa}))
\begin{eqnarray}
\label{eq-dnl-hf}
 D_{nl;nl}^{\text{HF}} [v_{\text{x}}] & \equiv  &
\langle \phi_{nlm}^{\text{HF}} | v_{\text{x}} -
\hat{v}_{\text{x}}\left[ \{\phi_a^{\text{HF}} \}\right]   | \phi_{nlm}^{\text{HF}}   \rangle =
\int_0^{\infty}dr \, \chi_{nl}^{\text{HF}}(r)
\left[  v_{\text{x}}(r) \chi_{nl}^{\text{HF}}(r) - F_{\text{x;}nl}^{\text{HF}}(r) \right] =  \nonumber \\
&&   C_{nl} -  \int_0^{\infty}dr \, \chi_{nl}^{\text{HF}}(r) U_{nl}(r)  \, ,
\end{eqnarray}
and, consequently, to express  $W_{nl}^{\perp;\text{rad}}$, Eq. (\ref{eq-wnl}),
as a linear functional of $U_{nl}$, namely
\begin{eqnarray}
\label{eq-wnl-unl}
W_{nl}^{\perp;\text{rad}}[v_{\text{x}},\{\phi_a^{\text{HF}}\}] (r)  =
U_{nl}(r) -  \chi_{nl}^{\text{HF}}(r) \int_0^{\infty}dr' \, \chi_{nl}^{\text{HF}}(r') U_{nl}(r') \, .
\end{eqnarray}
Thus, the terms $W_{nl}^{\perp;\text{rad}}[v_{\text{x}},\{\phi_a^{\text{HF}}\}]$ are small
for any potential $v_{\text{x}}$ that gives small $U_{nl}$,
and we also get  $D_{H,H}^{\text{HF}} \left[ v_{\text{x}}\right] \approx 0$, due to $C_H=0$,
from Eq.  (\ref{eq-dnl-hf}).
This means that such a potential $v_{\text{x}}$ belongs to  ${\cal V}_0$.
Obviously, the appropriate constants
$C_{nl }=\widetilde{C}_{nl }[v_{\text{x}}]$ that yield small $U_{nl}(r)$ for  $v_{\text{x}} \in {\cal V}_0$
are not strictly (and, thus, not uniquely) defined with this requirement.
However, according to Eq.  (\ref{eq-dnl-hf}),
satisfactory values of
$C_{nl }=\widetilde{C}_{nl }[v_{\text{x}}]$
are close to $D_{nl;nl}^{\text{HF}} [v_{\text{x}}]$, i.e.,
\begin{eqnarray}
\label{eq-dnl-hf2}
 \widetilde{C}_{nl }[v_{\text{x}}] \approx D_{nl;nl}^{\text{HF}} [v_{\text{x}}]  \, .
\end{eqnarray}
Note that the small, exponentially decaying, values of $U_{nl}(r)$ are obtained in the
asymptotic region for any non-diverging potentials $v_{\text{x}}$, especially for those
with the required, $-1/r$,  dependence for large $r$.

Each approximate exchange potential $v_{\text{x}} \in {\cal V}_0$  leads  to the KS orbitals
$\phi_a=\phi_a[v_{\text{x}}]$ (cf. Ref. \cite{note-etot-vx})
that are almost identical to the HF orbitals $\phi_a^{\text{HF}}$.
This can be shown by applying the perturbation-theory argument, presented
in Sec. \ref{sec-theory}, to the HF equation.
The orbital differences
$\Delta \phi_a^{\text{HF}} \equiv \phi_a[v_{\text{x}}]-\phi_a^{\text{HF}}$ are approximated
by the first-order corrections
$\delta \phi_a^{\text{HF}}({\bf r})=r^{-1}\delta \chi_{nl}^{\text{HF}}(r)Y_{lm}(\theta,\varphi)$ (where $a=(nlm)$)
which are given by
the equations (\ref{eq-dphi}), (\ref{eq-cta}) where the KS orbitals
$\phi_a$ and energies $\epsilon_a$ are replaced with
$\phi_a^{\text{HF}}$
 and $\epsilon_a^{\text{HF}}$, respectively, while
the perturbation $\Delta \hat{h}_{\text{HF}}$ is used instead of $\delta \hat{h}_{\text{s}}$.
This perturbation is given by the difference $\hat{h}_{\text{s}}-\hat{h}_{\text{HF}}$ of the one-body Hamiltonians
entering the KS and HF equations, Eqs. (\ref{eq-ks}), (\ref{eq-hf}),  correspondingly,
so that  it is the negative of the perturbation $\Delta \hat{h}_{\text{s}}$ considered in Sec. \ref{sec-vxoep-os}.
Presently, we write  $\Delta \hat{h}_{\text{HF}}=-\Delta \hat{h}_{\text{s}}$
in the following (selfconsistent) form
\begin{equation}
\label{eq-hf-per}
    \Delta \hat{h}_{\text{HF}}  =
    v_{\text{x}}-\hat{v}_{\text{x}}^{\text{F}}[\{\phi_a^{\text{HF}}\}] - \Delta v_{\text{es}} \,
\end{equation}
and we note that  the term
$\Delta v_{\text{es}}= v_{\text{es}}[n_{\text{tot}}^{\text{HF}}] - v_{\text{es}}[n_{\text{tot}}]$
is linear in  $\Delta \phi_a^{\text{HF}} \approx \delta \phi_a^{\text{HF}}$
in the leading order.
As a result, the equation (\ref{eq-dphi}) for $\delta \phi_a^{\text{HF}}$
leads to
a set of non-homogenous linear integral equations for
the corrections $\delta \chi_{nl}^{\text{HF}}$ to the HF occupied orbitals (of both spins).
In these equations,
the inhomogeneous terms (the right-hand sides) depend linearly on
$W_{nl}^{\perp;\text{rad}}[v_{\text{x}},\{\phi_a^{\text{HF}}\}]$,
through the matrix elements
\begin{eqnarray}
\label{eq-hf-per-inhomogen}
D_{n'l;nl}^{\text{HF}} [v_{\text{x}}]  \equiv
\langle \phi_{n'lm}^{\text{HF}} | v_{\text{x}}-\hat{v}_{\text{x}}^{\text{F}} [\{\phi_a^{\text{HF}}\}]
| \phi_{nlm}^{\text{HF}} \rangle=
 -   \int_0^{\infty} dr \chi_{n'l}^{\text{HF}}(r)
W_{nl}^{\perp;\text{rad}}[v_{\text{x}},\{\phi_a^{\text{HF}}\}](r) \, , \;\;
\end{eqnarray}
$n'\neq n$.
By solving the set of equations for $\delta \chi_{nl}^{\text{HF}}$, we can find
the radial orbital differences
$\chi_{nl}[v_{\text{x}}]-\chi_{nl}^{\text{HF}} \approx \delta \chi_{nl}^{\text{HF}}$,
which are small when all terms $W_{n'l'}^{\perp;\text{rad}}[v_{\text{x}},\{\phi_a^{\text{HF}}\}]$
are sufficiently small.
Now,  it can be claimed again  (cf. Sec. III.A) that, formally, it is  the norms  $\|\delta \phi_{nl}^{\text{HF}}\|$
that are the  adequate measure of smallness of $W_{nl}^{\perp;\text{rad}}[v_{\text{x}},\{\phi_a^{\text{HF}}\}]$.
Then, the class ${\cal V}_0={\cal V}_0(\eta)$ can be defined more precisely
with the condition $\| \phi_a [v_{\text{x}}] -\phi_a^{\text{HF}}\|\leq \eta$
(for $a=1,\ldots,N$) where $\eta \ll 1$.

The total energy
\begin{equation}\
\label{eq-en}
E\left [v_{\text{x}}\right]  \equiv
\langle \Psi[v_{\text{x}}]|\hat{H}|\Psi[v_{\text{x}}]\rangle =
E\left[ \{ \phi_a[v_{\text{x}}]\} \right] \, ,
\end{equation}
where $\Psi[v_{\text{x}}]$ is the Slater determinant constructed of $\{ \phi_a[v_{\text{x}}]\}$
(cf. Ref. \onlinecite{note-etot-vx}),  is very close to
$E_{\text{HF}}=E\left[ \{ \phi_a^{\text{HF}}\} \right] $  for any $v_{\text{x}} \in {\cal V}_0$
due to the orbital proximity, $\phi_a[v_{\text{x}}]\approx \phi_a^{\text{HF}}$.
As a result, the energies  $E[v_{\text{x}}]$, $v_{\text{x}} \in {\cal V}_0$,
 are also very close to $E[v_{\text{x}}^{\text{OEP}}]$
since the potential
$v_{\text{x}}^{\text{OEP}}$ minimizes the functional
$E[v_{\text{x}}] > E_{\text{HF}}$.
The obtained relation
$E[v_{\text{x}}] \approx E[v_{\text{x}}^{\text{OEP}}]$ 
implies, by the continuity of
the functional $E[v_{\text{x}}]$ (cf. Ref. \onlinecite{note-etot-vx}),
that the potentials $v_{\text{x}}$ belonging to ${\cal V}_0$ are close to
$v_{\text{x}}^{\text{OEP}}$ and, consequently, they are all close to each other.
Simultaneously, this argument  explains in a plausible way why the exact
exchange potential $v_{\text{x}}^{\text{OEP}}$ itself belongs to the class ${\cal V}_0$
and, in consequence, it gives the KS orbitals $\phi_a$
very close to $\phi_a^{\text{HF}}$.

Low magnitude of  $U_{nl}(r)$ obtained for a potential
$v_{\text{x}} \in {\cal V}_0$ implies that, {\em within} each occupied shell $S_n$,
the  shifted HF potentials
\begin{equation}
\label{eq-vxnl-hf}
\tilde{v}_{\text{x;}nl}^{\text{HF}}(r)=v_{\text{x;}nl}^{\text{HF}}(r)+C_{nl}
\end{equation}
($l\in {\cal L}_n \equiv \{0,\ldots,l_{\text{max}}^{(n)}\}$)
lie very close to $v_{\text{x}}(r)$,
\begin{equation}
\label{eq-hf-vxnl_vx}
\tilde{v}_{\text{x;}nl}^{\text{HF}}(r)  \approx v_{\text{x}}(r)  \; ,  \;\;
l \in {\cal L}_n \; , r\in S_n \, ,
\end{equation}
and, as a result, they
almost {\em coincide with  each other},
\begin{equation}
\label{eq-hf-rel-vxnl}
\tilde{v}_{\text{x;}nl}^{\text{HF}}(r) \approx \tilde{v}_{\text{x;}nl'}^{\text{HF}}(r) \; ,  \;\;
l, l' \in {\cal L}_n \; , r\in S_n \, .
\end{equation}
Similar proximity holds for the OEP potentials $\tilde{v}_{\text{x;}nl}(r)$,
since they are all very close to $v_{\text{x}}^{\text{OEP}}(r)$ within their respective shells
$S_n$; see Sec. \ref{sec-results-proximity} and Figs. \ref{fig-Ar_vx_vxa}, \ref{fig-Zn_vx_vxa}.

Let us note here that since any two different exchange potentials,
$v_{\text{x}}(r)$ and $v'_{\text{x}}(r)$,
from the class ${\cal V}_0$  are close to each other,
the respective constants,  $C_{nl}=\widetilde{C}_{nl}[v_{\text{x}}]$ and
$C_{nl}=\widetilde{C}_{nl}[v'_{\text{x}}]$,
that lead to small terms $U_{nl}(r)$, Eq. (\ref{eq-unl}),
are also close to each other. This is so because the equation
(\ref{eq-hf-vxnl_vx}) is satisfied for both potentials $v_{\text{x}}(r)$ and $v'_{\text{x}}(r)$,
as well as for each $(nl)$;  the same conclusion is reached by noting
that, with Eq. (\ref{eq-dnl-hf2}), we obtain the expression
\begin{equation}
\label{eq--cnl-vx-vxprim}
 \widetilde{C}_{nl}[v_{\text{x}}]-\widetilde{C}_{nl}[v'_{\text{x}}]  \approx
 D_{nl}^{\text{HF}}[v_{\text{x}}]-D_{nl}^{\text{HF}}[v'_{\text{x}}]=
 \int_0^{\infty} dr \left[ v_{\text{x}}(r)-v'_{\text{x}}(r)\right] \left[\chi_{nl}^{\text{HF}}(r)\right]^2 \, .
\end{equation}
which is small for $v_{\text{x}} \approx v'_{\text{x}}$.
In particular, by taking $v'_{\text{x}}=v_{\text{x}}^{\text{OEP}}$ we find
\begin{equation}
\label{eq-cnl-dnloep}
\widetilde{C}_{nl}[v_{\text{x}}] \approx \widetilde{C}_{nl}[v_{\text{x}}^{\text{OEP}}] \approx
D_{nl}^{\text{HF}}[v_{\text{x}}^{\text{OEP}}] \approx
D_{nl}[v_{\text{x}}^{\text{OEP}}]
\end{equation}
(for $v_{\text{x}} \in {\cal V}_0$) where the relation (\ref{eq-dnl-hf2}) and the orbital proximity,
$\phi_a[v_{\text{x}}^{\text{OEP}}] \approx \phi_a^{\text{HF}}$ are also  applied.
This means that the quantities $D_{nl}[v_{\text{x}}^{\text{OEP}}]$, found with
the exact exchange potential $v_{\text{x}}^{\text{OEP}}$, can be used  as the constants
$C_{nl}=\widetilde{C}_{nl}[v_{\text{x}}] $ suitable for all $v_{\text{x}} \in {\cal V}_0$.
Another possible set of  constants $\{C_{nl}\}$,
which can be determined easier
than  $D_{nl}[v_{\text{x}}^{\text{OEP}}]$, is given by the quantities
$\{C_{nl}^{\text{KLI-HF}}\}=D_{nl}^{\text{HF}}[v_{\text{x}}^{\text{KLI-HF}}]$  found, in a selfconsistent way,
for the KLI potential
$v_{\text{x}}^{\text{KLI-HF}}=v_{\text{x}}^{\text{KLI}}[\{\phi_a^{\text{HF}}\}]$
obtained with HF orbitals;
see Sec. \ref{sec-results-represent-vx-kli}, \ref{sec-results-represent-comparison} below.
The two sets of constants, listed in Table \ref{tab-dnl}, are indeed very close to each other.

A generalization of Eq.
(\ref{eq-hf-rel-vxnl}) is found when, in the expression (\ref{eq-unl}) for
$U_{nl}(r)$, the potential $v_{\text{x}}(r)$ is replaced
by $\tilde{v}_{\text{x;}n'l'}^{\text{HF}}(r)$ for $r \in S_{n'}$ according to  Eq. (\ref{eq-hf-vxnl_vx}),
the definition (\ref{eq-vxnl-hf}) is used, and
the smallness of $U_{nl}(r)$ for $v_{\text{x}} \in {\cal V}_0$ is accounted for.
The generalized relation reads
\begin{equation}
\label{eq-hf-rel}
F_{\text{x;}nl}^{\text{HF}}(r)+C_{nl}\chi_{nl}^{\text{HF}}(r) \approx
\left(v_{\text{x;}n'l'}^{\text{HF}}(r)+C_{n'l'} \right) \chi_{nl}^{\text{HF}}(r)\; ,\;r\in S_{n'}
\end{equation}
and it is satisfied for suitable set of constants $\{C_{nl}\}$ and for
all indices $(nl)$, $(n'l')$ corresponding to the occupied HF
orbitals, as well as for an appropriately chosen set of the shell border points
$r_{n,n+1}$.
The relation (\ref{eq-hf-rel}) is an {\em intrinsic} property of the HF orbitals
(and the Fock operator), since it is {\em not} implied by the DFT or
the definition of  $v_{\text{x}}^{\text{OEP}}$, though it has been
revealed here by inspecting the KS results for
$v_{\text{x}}=v_{\text{x}}^{\text{OEP}}$.
Obviously, the relation (\ref{eq-hf-rel}) can be rewritten as
\begin{equation}
\label{eq-hf-rel3}
\left(v_{\text{x;}nl}^{\text{HF}}(r)+C_{nl} \right) \chi_{nl}^{\text{HF}}(r) \approx
\left(v_{\text{x;}n'l'}^{\text{HF}}(r)+C_{n'l'} \right) \chi_{nl}^{\text{HF}}(r)\; ,\;r\in S_{n'}
\end{equation}
so that
by dividing  its both sides 
by $\chi_{nl}^{\text{HF}}(r)$ for $n'=n$,  we recover the approximate
equality (\ref{eq-hf-rel-vxnl}) of the shifted HF orbital exchange potentials  within the shell $S_n$.

The potentials $v_{\text{x;}nl}^{\text{HF}}(r)$ obtained with the occupied HF
orbitals $\chi_{nl}^{\text{HF}}(r)$ from
Eqs. (\ref{eq-fxnl}), (\ref{eq-vxnl})  do not diverge for $r\rightarrow \infty$,
unlike the KS potentials $v_{\text{x;}nl}(r)$, except for the HOMO one
(as well as $v_{\text{x;}1s}(r)$  in  the Be atom).
It is due to the form of the large-$r$ dependence
$\chi_{nl}^{\text{HF}}(r) \sim r^{\alpha_{nl}} e^{-\beta_H r}$
with a {\em common} (for all $nl$) coefficient $\beta_H=\sqrt{-2\epsilon_{H}^{\text{HF}}}$
of the exponential decay and with  the  appropriate values of
the orbital-specific constants $\alpha_{nl}$, cf. Refs.  \onlinecite{HMS69,HSS80,DJM84,IO92}.
As a result, the following asymptotic dependence for the HF exchange potentials is obtained
\begin{equation}
\label{eq-vxnl-hf-asymp}
v_{\text{x;}nl}^{\text{HF}}(r) =
-\left(\epsilon_{H}^{\text{HF}}-\epsilon_{nl}^{\text{HF}}\right) +
\frac{q_{nl}}{r} + o\left( \frac{1}{r}\right)
\end{equation}
where the constants $q_{nl}$ can differ from -1; see Appendix.
This dependence is confirmed by the numerical results obtained for the HF orbitals found
by solving  the HF equation with the highly accurate pseudospectral method \cite{MC09-prep};
see Fig. \ref{fig-Be_Ar-asymp}.
The constant term  in Eq. (\ref{eq-vxnl-hf-asymp}) vanishes only
for the HOMO potential $v_{\text{x;}H}^{\text{HF}}(r)$
and, in this case, we also find $q_{H}=-1$.
Thus, the exchange potential
$v_{\text{x;}H}^{\text{HF}}(r)$
has the $-1/r$ dependence for large $r$.
In consequence,
it is close to the potentials $v_{\text{x}} \in {\cal V}_0$
not only within the region $S_n$ of the shell
to which the HOMO belongs, but also in the asymptotic region $S_{\infty}$ where
these potentials  decay like $-1/r$.
The asymptotic dependence
\begin{equation}
\label{eq-hf-rel-asymp}
v_{\text{x;}H}^{\text{HF}}(r) = -\frac{1}{r} + o\left( \frac{1}{r}\right)\, , \;\;\; r\in S_{\infty} \, ,
\end{equation}
complements the relations (\ref{eq-hf-rel-vxnl}), (\ref{eq-hf-rel}), valid within the occupied shells $S_n$.
Note that the shifted potential $\tilde{v}_{\text{x;}H}^{\text{HF}}(r)$, entering Eq. (\ref{eq-hf-vxnl_vx}),
is equal to $v_{\text{x;}H}^{\text{HF}}(r)$ since we set
$C_{H}=0$ as in the definition of the class ${\cal V}_0$.

\subsection{Accurate representations of exact exchange potential with HF orbital exchange potentials}
\label{sec-results-represent-vx}

It has been shown above that the proximity of the individual HF and exchange-only KS-OEP occupied
orbitals implies
the relations (\ref{eq-hf-rel-vxnl}), (\ref{eq-hf-rel}) satisfied by the HF orbitals.
Interestingly, the converse is also true.
Namely, assuming that the relation (\ref{eq-hf-rel}) holds
(then, the relation (\ref{eq-hf-rel-vxnl}) is also true) and the constants
$C_{nl}$ which satisfy it  are known, we can
effectively construct local exchange potentials
$v_{\text{x}}(r)$ that belong to the class ${\cal V}_0$, i.e.,
which lead to small terms $U_{nl}(r)$, have correct ($-1/r$) asymptotic behaviour,
and, in consequence, give the KS orbitals close to the HF ones.
As it is argued above, such potentials should represent the exact exchange
potential $v_{\text{x}}^{\text{OEP}}(r)$ with high accuracy.

\subsubsection{Shell-resolved piecewise exchange potentials}
\label{sec-results-represent-vx-shell-resolved}

If the relations  (\ref{eq-hf-rel-vxnl}), (\ref{eq-hf-rel}) are fulfilled for
a given set of the constants $C_{nl}$,
the straightforward way to build a potential $v_{\text{x}} \in {\cal V}_0$
is to set it equal to  one of the (almost coinciding) potentials   $\tilde{v}_{\text{x;}nl}^{\text{HF}}(r)$,
Eq. (\ref{eq-vxnl-hf}), in each occupied atomic shell $S_n$;
then,  the resulting potential $v_{\text{x}}$ satisfies the relation (\ref{eq-hf-vxnl_vx})
(which has to hold for any  $v_{\text{x}} \in {\cal V}_0$).
%
%
In particular, we can choose the $s$-orbital ($l=0$) potentials $\tilde{v}_{\text{x;}n0}^{\text{HF}}(r)$
for $r \in S_n$, $n=1,\ldots,n_{\text{occ}}$.
However, within the outmost occupied  shell $S_n$, $n=n_{\text{occ}}$,
it is better to use the HOMO exchange potential $v_{\text{x;}H}^{\text{HF}}(r)$
since it can represent the constructed $v_{\text{x}}(r)$ not only
for $r \in S_n$, $n=n_{\text{occ}}$, but also in the asymptotic region $S_{\infty}$ where
it has the $-1/r$ decay (required for $v_{\text{x}} \in {\cal V}_0$), cf. Eq. (\ref{eq-hf-rel-asymp}).
In this way, a  {\em piecewise} (pw)  exchange potential
\begin{equation}
\label{vx-pw0}
v_{\text{x}}^{\text{pw,0}}(r) = \sum_{n=1}^{n_{\text{occ}}-1}\theta_n^{\text{HF,0}}(r) \tilde{v}_{\text{x;}n0}^{\text{HF}}(r)
+ \theta_{n_{\text{occ}}}^{\text{HF,0}}(r) v_{\text{x;}H}^{\text{HF}}(r) \, ,
\end{equation}
is obtained; here  the step-like functions $\theta_n^{\text{HF,0}}(r)$ are equal to
\begin{subequations}
\label{eq-theta0}
\begin{eqnarray}
&&\theta(r-r_{n-1,n}^{\text{HF,0}})\theta(r_{n,n+1}^{\text{HF,0}}-r) \hspace{1cm}
 (n<n_{\text{occ}})\, , \\
&&\theta(r-r_{n,n-1}^{\text{HF,0}})
\hspace{3.3cm}(n=n_{\text{occ}})\, .
\end{eqnarray}
\end{subequations}
This construction is restricted to the case when  the HOMO belongs
to the outmost occupied shell, which is true for  the closed-$l$-shell atoms.

To make the potential $v_{\text{x}}^{\text{pw,0}}(r) $ {\em continuous},
the shell borders  $r_{n,n+1}$
are set
at the points $r_{n,n+1}^{\text{HF,0}}$, $n=1,\ldots,n_{\text{occ}}-1$,
where its constituent potentials
from the neighboring shells, $S_n$ and $S_{n+1}$,  match, i.e., the condition
\begin{subequations}
\label{eq-rnn1-hf0-cond}
\begin{eqnarray}
\tilde{v}_{\text{x;}n0}^{\text{HF}}(r) & = &\tilde{v}_{\text{x;}n+1,0}^{\text{HF}}(r)  \hspace{1cm}
(n \leq n_{\text{occ}}-2 ) \, , \\
\tilde{v}_{\text{x;}n0}^{\text{HF}}(r) & = & v_{\text{x;}H}^{\text{HF}}(r) \hspace{1.6cm}  (n=n_{\text{occ}}-1) \,
\end{eqnarray}
\end{subequations}
is satisfied for $r=r_{n,n+1}^{\text{HF,0}}$;  we also define $r_{01}^{\text{HF,0}}=0$.
The outer border $r_{n,n+1}$ of the outmost occupied shell $S_n$, $n=n_{\text{occ}}$,
does not have to be defined since  it is not used in Eqs. (\ref{vx-pw0}-\ref{eq-theta0}).
However,  if the point  $r_{n,n+1}$, $n=n_{\text{occ}}$, needs to be determined
(e.g., when we want to specify  the region $S_n$ where the relations
(\ref{eq-hf-rel-vxnl}), (\ref{eq-hf-rel}) or (\ref{eq-hf-vxnl_vx})  are fulfilled for $n=n_{\text{occ}}$)
it can be plausibly defined as the smallest of
the classical turning  points $r_{nl}^{\text{TP}}$ for electrons from the $n_{\text{occ}}$-th shell;
in the HF case, each point $r_{nl}^{\text{TP}}$ can be found from the condition
$v_{\text{s;}nl}^{\text{HF}}(r)+l(l+1)/(2r^2)=\epsilon_{nl}^{\text{HF}}$; cf. Eq. (\ref{eq-vsa-hf}).

The relation (\ref{eq-hf-rel})
(with $l'=l_{\text{max}}^{(n')}$  for $n'= n_{\text{occ}}$ (HOMO), and  $l'=0$
for $n'\leq n_{\text{occ}}-1$ )
and Eq. (\ref{eq-hf-rel-asymp})  immediately imply that the constructed potential
$v_{\text{x}}(r)=v_{\text{x}}^{\text{pw,0}}(r)$ yields small $U_{nl}(r)$, Eq. (\ref{eq-unl}),
within each occupied shell $S_{n'}$ and also for  $r \in S_{\infty}$.
Thus, the potential $v_{\text{x}}^{\text{pw,0}}(r)$ belongs to the class ${\cal V}_0$ and, in consequence,
it is close to $v_{\text{x}}^{\text{OEP}}(r)$; this
conclusion is  supported by the numerical results plotted in Fig. \ref{fig-Ar_vxapp-vxoep}(a).
Such numerical confirmation also implies that the points $r_{n,n+1}^{\text{HF,0}}$
can  be indeed be  chosen  for use as the shell borders $r_{n,n+1}$ in the relations
(\ref{eq-hf-rel}), (\ref{eq-hf-rel-asymp}).

Another  representation  of $v_{\text{x}}^{\text{OEP}}(r)$
is obtained by constructing
a continuous piecewise potential
\begin{equation}
\label{vx-pw}
v_{\text{x}}^{\text{pw}}(r) = \sum_{n=1}^{n_{\text{occ}}}\theta_n^{\text{HF}}(r) v_{\text{x}}^{(n)}(r)
\end{equation}
formed from the HF {\em shell} exchange potentials
\begin{equation}
\label{vx-shell}
 v_{\text{x}}^{(n)}(r) \equiv
 \sum_{l\in{\cal L}_n}
 \tilde{v}_{\text{x;}nl}^{\text{HF}}(r) \frac{\rho_{nl}^{\text{HF}}(r)}{ \rho_n^{\text{HF}}(r)}\, ,
\end{equation}
each applied in its shell region $S_n$. The points
$r_{n,n+1}^{\text{HF}}$ defining the shell borders
are now the solutions of the continuity equation
\begin{equation}
\label{eq-rnn1-hf-cond}
v_{\text{x}}^{(n)}(r)=v_{\text{x}}^{(n+1)}(r)
\end{equation}
for
$n=1,2,\ldots,n_{\text{occ}}-1$; $r_{0,1}^{\text{HF}}=0$. We denote
\begin{subequations}
\begin{eqnarray}
\label{rho-nl-hf}
\rho_{nl}^{\text{HF}}(r) &=& (2l+1)\left[\chi_{nl}^{\text{HF}}(r)\right]^2 \,,\\
\label{rho-n-hf}
\rho_n^{\text{HF}}(r) &=&\sum_{l\in{\cal L}_n} \rho_{nl}^{\text{HF}}(r) \, ,\\
\label{rh-hf}
  \rho^{\text{HF}}(r) &=& \sum_{n=1}^{n_{\text{occ}}} \rho_{n}^{\text{HF}}(r) \, ,
\end{eqnarray}
\end{subequations}
and the functions $\theta_n^{\text{HF}}(r)$ are defined like $\theta_n^{\text{HF,0}}(r)$
with Eq. (\ref{eq-theta0})
where the radii $r_{n,n+1}^{\text{HF,0}}$ are replaced by $r_{n,n+1}^{\text{HF}}$.
Each shell potential $v_{\text{x}}^{(n)}(r)$ is very close to the
almost coinciding potentials
$\tilde{v}_{\text{x;}nl}^{\text{HF}}(r)$,
$l \in {\cal L}_n$,
for $r\in S_n$, Eq. (\ref{eq-hf-rel-vxnl}).
Thus, the potential $v_{\text{x}}^{(n')}$ can be substituted for $\tilde{v}_{\text{x;}n'l'}^{\text{HF}}$
in Eq. (\ref{eq-hf-rel}),  which leads to the relation
\begin{equation}
\label{eq-hf-rel2}
F_{\text{x;}nl}^{\text{HF}}(r)+C_{nl}\chi_{nl}^{\text{HF}}(r) \approx
v_{\text{x}}^{(n')}(r) \chi_{nl}^{\text{HF}}(r)\, ,
\end{equation}
valid for $r\in S_{n'}$ and  $n'=1,\ldots,n_{\text{occ}}$.
It means that the terms $U_{nl}(r)$, Eq. (\ref{eq-unl}),  are small
for  the potential $v_{\text{x}}(r)=v_{\text{x}}^{\text{pw}}(r)$ within each occupied shell $S_{n'}$.
This implies that the potential $v_{\text{x}}^{\text{pw}}$
belongs to the class ${\cal V}_0$ and, hence, it is close to $v_{\text{x}}^{\text{OEP}}$;
 cf. Fig. \ref{fig-Ar_vxapp-vxoep}(b).
To make this argument complete we note that
the potential $v_{\text{x}}^{\text{pw}}(r)$ is also close to $v_{\text{x;}H}^{\text{HF}}(r)$
in the asymptotic region $S_{\infty}$ and, thus it has the correct, $-1/r$, dependence
for $r \in S_{\infty}$ (which is a property requested for potentials $v_{\text{x}} \in {\cal V}_0$).
Indeed, for large $r$,
the factor $\rho_{nl}^{\text{HF}}(r) / \rho_n^{\text{HF}}(r)$,
present in Eq. (\ref{vx-shell}), goes to 1 for $nl=H$ and it vanishes
like $r^{-q}$, $q\geq 4$, for other HF occupied orbitals $\chi_{nl}^{\text{HF}}$;
(see Eqs. (\ref{eq-chi-hf-asymp}), (\ref{eq-chi-hf-asymp2}) in Appendix;
cf. Refs. \onlinecite{HMS69,HSS80,DJM84,IO92}).
The presented construction of $v_{\text{x}}^{\text{pw}}(r)$
is restricted to the case of the closed-$l$-shell atoms where the HOMO belongs
to the outmost occupied shell.

\subsubsection{Shell-dependent  slope of the DFT exchange potential}
\label{sec-results-represent-vx-shell-slope}

The slope of the  exact exchange  potential $v_{\text{x}}^{\text{OEP}}(r)$
changes rather abruptly (here disregarding small intershell bumps)
when we move through an atom, from  one atomic shell to the next one;
cf. Fig. \ref{fig-Be_vx_vxa}, \ref{fig-Ar_vx_vxa}, \ref{fig-Zn_vx_vxa}.
This property can be explained by the fact that the potential $v_{\text{x}}^{\text{OEP}}(r)$
is represented with high accuracy, within each occupied shell $S_n$,
by the potentials $\tilde{v}_{\text{x;}nl}^{\text{HF}}(r)$, $l \in {\cal L}_n$, and,
in particular,  by the $s$-orbital exchange potentials
$\tilde{v}_{\text{x;}n0}^{\text{HF}}(r)$
which exist for each occupied shell ($n=1,\ldots,n_{\text{occ}}$).
Indeed, the slope $d \tilde{v}_{\text{x;}n0}^{\text{HF}} /dr=d v_{\text{x;}n0}^{\text{HF}} /dr$ found
within the shell $S_n$    for  the potential
$\tilde{v}_{\text{x;}n0}^{\text{HF}}(r)=v_{\text{x;}n0}^{\text{HF}}(r)+C_{n0}$ obtained with
the Eqs.  (\ref{eq-fxnl}), (\ref{eq-vxnl}) (where the orbitals $\{\chi_{nl}\}$ are replaced
by  $\{\chi_{nl}^{\text{HF}}\}$),
is distinctively different from the slopes of
other potentials $\tilde{v}_{\text{x;}n'0}^{\text{HF}}(r)$ within their respective shells $S_{n'}$.
It is related to the fact that the orbitals $v_{\text{x;}nl}^{\text{HF}}(r)$
(e.g., for $l=0$)  corresponding to different atomic shells
are  localized at different distances from the nucleus.

The above general argument readily applies to the Be atom.
In this case, the potential
\begin{subequations}
\begin{equation}
\label{eq-v1s-hf-be}
v_{\text{x;}1s}^{\text{HF}}(r)=v_0^{\text{HF}}(1s,1s;r)+\kappa_{2s,1s}(r)v_0^{\text{HF}}(2s,1s;r)
\end{equation}
(cf. Eq. (\ref{eq-v1s-be}))
is  very well represented, for $r \in  S_1$,
by the first term $v_0(1s,1s;r)$; see Fig. \ref{fig_Be_vxnl_part}.
The other term in Eq. (\ref{eq-v1s-hf-be})
is  much smaller
due to the combined effect of the small ratio
$\kappa_{2s,1s}(r)=\chi_{2s}^{\text{HF}}(r)/\chi_{1s}^{\text{HF}}(r)$ (we find
$|\kappa_{2s,1s}(r)| <0.2$ for $0\leq r\leq 0.74\,\text{a.u.}$)
and low magnitude of
$v_0^{\text{HF}}(2s,1s;r)=\int_r^{\infty} dr' (1/r-1/r') \chi_{1s}^{\text{HF}}(r')\chi_{2s}^{\text{HF}}(r')$
(in comparison to $v_0^{\text{HF}}(1s,1s;r)$), which decreases with increasing $r$.
We also find that,
in the expression
\begin{equation}
\label{eq-v2s-hf-be}
v_{\text{x;}2s}^{\text{HF}}(r)=v_0^{\text{HF}}(2s,2s;r)+\kappa_{1s,2s}(r)v_0^{\text{HF}}(1s,2s;r)
\end{equation}
\end{subequations}
(cf. Eq. (\ref{eq-v2s-be})),
the term $v_0^{\text{HF}}(2s,2s;r)$ clearly dominates
within the shell $S_2$ where both the term $v_0^{\text{HF}}(1s,2s;r)$
and the ratio $\kappa_{1s,2s}(r)=\chi_{1s}^{\text{HF}}(r)/\chi_{2s}^{\text{HF}}(r)$
decay exponentially for the Be atom; cf.  Fig. \ref{fig_Be_vxnl_part}. Thus,  we obtain the relation
\begin{equation}
\label{eq-v1s2s-deriv}
    \frac{d v_{\text{x;}nl}^{\text{HF}}(r)}{dr}  \approx  \frac{d v_0^{\text{HF}}(nl,nl;r)}{dr}
    = -\frac{Q_{nl}(r)}{r^2} \, , \;\; r \in S_n
\end{equation}
which holds for both $nl=1s$ and $2s$; here $Q_{nl}(r)=\int_0^r dr' [\chi_{nl}^{\text{HF}}(r')]^2$.
As a result, we conclude that the derivative
$d v_{\text{x;}1s}^{\text{HF}}(r)/dr \approx -3.65\,\text{a.u.}$ at the point
$r=r_{1s}^{*}=  0.37 \,\text{a.u} \in S_1$ where $Q_{1s}(r)=0.5$
has the magnitude (approximately) $(r_{2s}^{*}/r_{1s}^{*})^2=44.2 $ times larger than the slope
  $d v_{\text{x;}2s}^{\text{HF}} (r)/dr  \approx -0.0826\,\text{a.u.}$
  at the point $r=r_{2s}^{*}=2.46 \,\text{a.u} \in S_2$
  where $Q_{2s}(r)=0.5$.
These estimates agree well with the values $-3.70\, \text{a.u}$  and $-0.0846\, \text{a.u}$ of
$dv_{\text{x}}^{\text{OEP}}(r)/dr$ at the points $r=r_{1s}^{*}$  and $r=r_{2s}^{*}$, respectively;
the ratio of these two slopes is 43.7.

\subsubsection{KLI-  and LHF(CEDA)-like potentials constructed from the HF orbitals}
\label{sec-results-represent-vx-kli}

The KLI-like potential  $\breve{v}_{\text{x}}^{\text{KLI}}(r)$ can be defined
for  the HF orbitals  $\{\phi_{nlm}^{\text{HF}}\}$ and the constants $\{C_{nl}\}$
by substituting them for $\{\phi_{nlm}\}$ and $\{D_{aa}\}=\{D_{nl;nl}\}$, respectively,
in Eqs. (\ref{eq-vx-kli-oep}-\ref{eq-vx-es}).
It takes the following form
\begin{eqnarray}
\label{vx-kli-hf-breve}
  \breve{v}_{\text{x}}^{\text{KLI-HF}}\left[ \{C_{nl}\} \right](r) \equiv
\breve{v}_{\text{x}}^{\text{KLI}}\left[ \{\chi_{nl}^{\text{HF}}\}, \{C_{nl}\} \right](r) =
 \nonumber \\
 \frac{1}{\rho^{\text{HF}}(r)} \sum_{nl}^{\text{occ}}  (2l+1)
\left[F_{\text{x;}nl}^{\text{HF}}(r)+C_{nl}\chi_{nl}^{\text{HF}}(r)\right]
 \chi_{nl}^{\text{HF}}(r)
\end{eqnarray}
where, for the closed-$l$-shell atoms, the quantities $\chi_{nl}^{\text{HF}}$ and $C_{nl}$
are indicated as the effective arguments of $\breve{v}_{\text{x}}^{\text{KLI}}$.
This potential can also be expressed in terms of the HF orbital  exchange potentials,
\begin{equation}
\label{vx-kli-hf-breve-vxnl}
 \breve{v}_{\text{x}}^{\text{KLI-HF}}\left[ \{C_{nl}\} \right](r)  =
 \sum_{nl}^{\text{occ}}
 \frac{\rho_{nl}^{\text{HF}}(r)}{\rho^{\text{HF}}(r)} \tilde{v}_{\text{x;}nl}^{\text{HF}}(r) \, .
\end{equation}
It can be argued that the  potential $\breve{v}_{\text{x}}^{\text{KLI-HF}}(r)$, Eq. (\ref{vx-kli-hf-breve}),
is close to
$\breve{v}_{\text{x}}^{\text{KLI}}[ \{\chi_{nl}^{\text{OEP}}\}, \{D_{nl;nl}^{\text{OEP}}\} ](r)$,
and, consequently,
also to $v_{\text{x}}^{\text{OEP}}(r)$ (cf. Sec. \ref{sec-vx-kli-lhf}), because the HF
orbitals $\chi_{nl}^{\text{HF}}(r)$ nearly coincide with  the KS-OEP orbitals
$\chi_{nl}^{\text{OEP}}(r)$, while
the constants $C_{nl}$ satisfying the relation (\ref{eq-hf-rel}) are very close to
$D_{nl;nl}^{\text{OEP}} \equiv D_{nl;nl}[v_{\text{x}}^{\text{OEP}}]$;  cf. Eq. (\ref{eq-cnl-dnloep}).
However, the high-quality of the KLI-like potential
$\breve{v}_{\text{x}}^{\text{KLI-HF}}(r)$
is, in fact, a {\em direct} consequence of the relation  (\ref{eq-hf-rel})
revealed for the HF orbitals.
Indeed, this relation immediately implies that the potential given by Eq. (\ref{vx-kli-hf-breve})
is  close to $v_{\text{x;}n'l'}^{\text{HF}}(r)$, $l' \in {\cal L}_{n'}$, 
for $r \in S_{n'}$, $n'=1,\ldots, n_{\text{occ}}$.
This means, in particular, that the  potential
$\breve{v}_{\text{x}}^{\text{KLI-HF}}(r)$
is close to $v_{\text{x}}^{\text{pw,0}}(r)$ within each occupied shell $S_{n'}$
so that it also yields small   terms $U_{nl}(r)$ there (for any $(nl)\in \text{occ}$).
For large $r$,  the potential $\breve{v}_{\text{x}}^{\text{KLI-HF}}(r)$, given by
Eq. (\ref{vx-kli-hf-breve-vxnl}) (with $C_{H}=0$), becomes close to
$v_{\text{x;}H}^{\text{HF}}(r)$
so that it decays like $-1/r$ (see the discussion  for $v_{\text{x}}^{\text{pw}}(r)$ above).
These properties of the potential  $\breve{v}_{\text{x}}^{\text{KLI-HF}}$ imply
that it belongs to the class $ {\cal V}_0$
and, in consequence, it is close to $v_{\text{x}}^{\text{OEP}}$,
cf. Fig. \ref{fig-Ar_vxapp-vxoep}(c).

In particular, this is true for the KLI potential
\begin{eqnarray}
\label{vx-kli-hf}
v_{\text{x}}^{\text{KLI-HF}}(r)
  \equiv  \breve{v}_{\text{x}}^{\text{KLI-HF}}\left[ \{C_{nl}^{\text{KLI-HF}}\} \right](r)
 =  v_{\text{x}}^{\text{KLI}}\left[ \{\chi_{nl}^{\text{HF}}\} \right ](r)
\end{eqnarray}
calculated with Eq. (\ref{eq-vx-kli-orig}) for the HF orbitals and the constants $C_{nl}^{\text{KLI-HF}}$
that are found from their self-consistency condition
\begin{equation}
\label{cond-dnl-kli}
D_{nl;nl}^{\text{HF}}[v_{\text{x}}^{\text{KLI-HF}}]=C_{nl}^{\text{KLI-HF}}
\end{equation}
given in Ref. \cite{KLI92}.
To show this, let us express them as the sum
\begin{equation}
\label{eq-kli-dcnl-def}
C_{nl}^{\text{KLI-HF}}=C_{nl}+\Delta C_{nl} \, ,
\end{equation}
where the constants $C_{nl}$  satisfy the relation (\ref{eq-hf-rel}).
Then, we obtain, from Eq. (\ref{eq-dnl-hf}) (the first line) and
Eqs. (\ref{vx-kli-hf-breve}), (\ref{cond-dnl-kli}),
the following set of linear equations for $\Delta C_{nl}$
\begin{eqnarray}
\label{eq-kli-dcnl-eqn}
 \Delta C_{nl}  - \sum_{n'l'}
(2l'+1) \int_0^{\infty} dr
\frac{\left[ \chi_{nl}^{\text{HF}}(r) \chi_{n'l'}^{\text{HF}}(r) \right]^2}
{ \rho^{\text{HF}}(r)}   \Delta C_{n'l'}
= -  \int_0^{\infty} dr \chi_{nl}^{\text{HF}}(r)  U_{nl} [\breve{v}_{\text{x}}^{\text{KLI-HF}}](r)
\end{eqnarray}
($n=1,\ldots,n_{\text{occ}}$, $l \in {\cal L}_n$; $nl\neq H$)
where the right-hand side includes the potential
$\breve{v}_{\text{x}}^{\text{KLI-HF}}=\breve{v}_{\text{x}}^{\text{KLI-HF}}[\{C_{nl}\}]$,
Eq. (\ref{vx-kli-hf-breve}).
The set of equations (\ref{eq-kli-dcnl-eqn}) remains  satisfied
when a common constant is added to each
$\Delta C_{nl}$. Therefore, to make this  set 
well-defined,  we put $\Delta C_{H}=0$  and, simultaneously,
exclude  the equation for $nl=H$ from the set
(then, we find $C_{H}^{\text{KLI-HF}}=C_{H}=0$).
Since the terms $U_{nl} [\breve{v}_{\text{x}}^{\text{KLI-HF}}]$
(calculated for $C_{nl}$ satisfying Eq. (\ref{eq-hf-rel}))
are small, the corrections $\Delta C_{nl}$  obtained
by solving  the equations (\ref{eq-kli-dcnl-eqn}) are also small.
This means that the potential $v_{\text{x}}^{\text{KLI-HF}}$ is very close to
$\breve{v}_{\text{x}}^{\text{KLI-HF}}[\{C_{nl}\}] \in {\cal V}_0$ and, in consequence,
this potential itself belongs to the class ${\cal V}_0$.

The KLI condition (\ref{cond-dnl-kli}) can also be satisfied by minimizing, with respect to the constants
$\{C_{nl}\}$, the function
\begin{eqnarray}
\label{min-fun-kli}
g\left( \{C_{nl}\} \right)   \equiv  \sum_{a}^{\text{occ}}
\| v_{\text{x}}\phi_a^{\text{HF}} - \hat{v}_{\text{x}}^{\text{F}}\phi_a^{\text{HF}} -
C_{nl}\phi_a^{\text{HF}} \|^2 =
 \sum_{nl}^{\text{occ}}  (2 l+1) \int_0^{\infty} dr \, \Bigl( U_{nl}[v_{\text{x}}](r) \Bigr)^2
\end{eqnarray}
where we put
$v_{\text{x}}=\breve{v}_{\text{x}}^{\text{KLI-HF}}[\{C_{nl}\}]$, Eq. (\ref{vx-kli-hf-breve}),
and  $a=(nlm)$;  a similar  expression 
leads, after minimization, to the selfconsistent constants $\{D_{ab}\}$ for
the LHF (CEDA) approximate potential \cite{HC05,SSD06b}.
To avoid the presence of an arbitrary common constant that can be  added to all
$C_{nl}=C_{nl}^{\text{KLI-HF}}$ (since such addition does not change the value of
$g\left( \{C_{nl}\} \right)$), we again set $C_{H}^{\text{KLI-HF}}=0$ in Eq. (\ref{min-fun-kli}).
The function $g\left( \{C_{nl}\} \right)$ attains very small value for the constants
$\{C_{nl}\}$ that satisfy the relation (\ref{eq-hf-rel}) since 
they lead to small terms $U_{nl}[v_{\text{x}}]$ for
$v_{\text{x}}=\breve{v}_{\text{x}}^{\text{KLI-HF}}[\{C_{nl}\}]$.
The set of constants $\{C_{nl}\}=\{C_{nl}^{\text{KLI-HF}}\}$ that
minimizes the function (\ref{min-fun-kli})
 have to yield even lower value of $g\left( \{C_{nl}\} \right)$, and, in consequence,
 they should also give small terms  $U_{nl}[\breve{v}_{\text{x}}^{\text{KLI-HF}}]$.
 Thus, we can conclude again that  the corresponding potential
 $v_{\text{x}}^{\text{KLI-HF}}$ belongs to  ${\cal V}_0$.

By  extending the arguments  presented above for the KLI potential
we can show that the high accuracy of the LHF (CEDA) approximation
 is also {\em directly} explained by the revealed properties of the HF orbital exchange potentials.
Let us first note that the potential
$\breve{v}_{\text{x}}^{\text{KLI}}\left[ \{\chi_{nl}^{\text{HF}}\}, \{C_{nl}\} \right ]$
is a special case of the LHF-like potential
\begin{eqnarray}
\label{eq-vx-lhf-hf}
   \breve{v}_{\text{x}}^{\text{LHF-HF}}\left[ \{\{C_{n'l,nl}\} \right ] (r) =
  \breve{v}_{\text{x}}^{\text{LHF}}\left[ \{\chi_{nl}^{\text{HF}}\}, \{C_{n'l,nl}\} \right ](r)=  \nonumber \\
   \frac{1}{\rho^{\text{HF}}(r)} \sum_{nl,n'l}^{\text{occ}}  (2l+1)
\left[\delta_{n'n}F_{\text{x;}nl}^{\text{HF}}(r)+C_{n'l,nl}\chi_{n'l}^{\text{HF}}(r)\right]
 \chi_{nl}^{\text{HF}}(r) \nonumber \\
\end{eqnarray}
calculated  with Eq. (\ref{eq-vx-lhf}) for the HF orbitals and the constants
$C_{n'l,nl}=\delta_{n'n} C_{nl}$.
We can now solve the LHF self-consistency condition \cite{DSG01,GB01}
\begin{equation}
\label{cond-lhf}
    D_{n'l,nl} \left[v_{\text{x}}^{\text{LHF-HF}}  \right]= C_{n'l,nl}^{\text{LHF-HF}} \, ,
\end{equation}
where
\begin{equation}
\label{vx-lhf-hf}
v_{\text{x}}^{\text{LHF-HF}} (r) \equiv
\breve{v}_{\text{x}}^{\text{LHF}}\left[ \{\chi_{nl}^{\text{HF}}\}, \{C_{n'l;nl}^{\text{LHF-HF}} \} \right](r) \, ,
\end{equation}
by expressing $C_{n'l;nl}^{\text{LHF-HF}}$ as $\delta_{n'n}C_{nl}+\Delta C_{n'l,nl}$. 
Then, the corrections $\Delta C_{n'l,nl}$ satisfy a set of linear algebraic equations
(similar to  Eq. (\ref{eq-kli-dcnl-eqn})) 
where
the right-hand sides are given by the integrals
$-\int_0^{\infty} dr \chi_{n'l}^{\text{HF}}(r)  U_{nl}(r)$; we also set
$\Delta C_{H,H}=0$. 
The terms $U_{nl}(r)$ are small since they are calculated here for
$v_{\text{x}}=\breve{v}_{\text{x}}^{\text{KLI-HF}}\left[\{C_{nl}\} \right ]$
obtained with the constants $C_{nl}$ satisfying   the relation (\ref{eq-hf-rel}).
Thus, the resulting corrections $\Delta C_{n'l,nl}$  are also small.
This implies that
the potential $v_{\text{x}}^{\text{LHF-HF}}(r)$ is close to
$\breve{v}_{\text{x}}^{\text{KLI-HF}}\left[ \{C_{nl}\} \right](r)$ and,
as a result, it also gives
small $U_{nl}(r)$. In effect, the LHF exchange potential $v_{\text{x}}^{\text{LHF-HF}}$,
obtained with the HF orbitals,
belongs to the class ${\cal V}_0$ and it is close to $v_{\text{x}}^{\text{OEP}}$.

\subsubsection{Comparison of different approximate representations of exact exchange potential}
\label{sec-results-represent-comparison}

The constants $\{C_{nl}^{\text{KLI-HF}}\}$ obtained with the KLI selfconsistent condition
(\ref{cond-dnl-kli}) have been shown to differ only by small corrections $\{\Delta C_{nl}\}$ from
from any set of constants $\{C_{nl}\}$  satisfying the relation (\ref{eq-hf-rel}).
This  property combined with Eq. (\ref{eq-cnl-dnloep}) explains the  small magnitudes of
$C_{nl}^{\text{KLI-HF}}-D_{nl;nl}[v_{\text{x}}^{\text{OEP}}]$, cf. Table \ref{tab-dnl}.
It also implies 
that the constants $\{C_{nl}^{\text{KLI-HF}}\}$ themselves satisfy
the relation (\ref{eq-hf-rel})
so that they can indeed be used 
in construction of the approximate potentials
discussed in Sec. \ref{sec-results-represent-vx}.
In particular, as it is already mentioned above (see Fig. \ref{fig-Ar_vxapp-vxoep}),
the potentials $v_{\text{x}}^{\text{pw,0}}(r)$, $v_{\text{x}}^{\text{pw}}(r)$,
built entirely with the HF orbitals
and the constants $C_{nl}=C_{nl}^{\text{KLI-HF}}$,
are found  to be very accurate representations of the exact
exchange potential   $v_{\text{x}}^{\text{OEP}}(r)$.
The obtained quality of its approximation is almost the same as for the potentials
$v_{\text{x}}^{\text{KLI-HF}}(r)$, Eq. (\ref{vx-kli-hf}),  and
$\breve{v}_{\text{x}}^{\text{KLI}}(r)$, Eq. (\ref{eq-vx-kli-oep}),
the latter of which is built of the KS-OEP orbitals and it is the dominant part
of $v_{\text{x}}^{\text{OEP}}(r)$, Eq. (\ref{eq-vx-oep}).
However, any of the four approximate potentials fails to reproduce well the characteristic
bumps of $v_{\text{x}}^{\text{OEP}}(r)$ at the shell borders; cf.
Figs.  \ref{fig-Be_vx_vxa},  \ref{fig-Ar_vx_vxa}, \ref{fig-Zn_vx_vxa}, \ref{fig-Ar_vxapp-vxoep}.
Thus, it is  the minor part of the exact exchange potential, namely, its OS term
$v_{\text{x}}^{\text{OS}}(r)$, Eq. (\ref{vx-os}), depending linearly on
$\delta\phi_a({\bf r})$ ($a=1,\ldots,N$),  that produces these local maxima of $v_{\text{x}}^{\text{OEP}}(r)$.
This means that the intershell bumps of $v_{\text{x}}^{\text{OEP}}(r)$ are the consequence of
the finite (though very small) differences
$\phi_a({\bf r})-\phi_a^{\text{HF}}({\bf r}) \approx \delta\phi_a({\bf r})$ between the KS and HF occupied orbitals.

The   potentials $v_{\text{x}}^{\text{pw,0}}(r)$ and
$v_{\text{x}}^{\text{pw}}(r)$ are expressed, in each atomic shell $S_n$,
in terms of   the  orbital exchange potentials
$\tilde{v}_{\text{x;}nl}^{\text{HF}}(r)$, $l \in {\cal L}_n$,
that correspond to this shell only.
This  feature
makes these two representations of the exact exchange potential 
be  significantly different from the KLI-like potential, Eq. (\ref{vx-kli-hf-breve-vxnl}).
Indeed, the latter  depends, within each shell $S_{n}$,  on all potentials
$\tilde{v}_{\text{x;}n'l}^{\text{HF}}(r)$, corresponding to both
the same  ($n'=n$)  and other ($n' \neq n$) shells.
In consequence, the KLI-like potential Eq. (\ref{vx-kli-hf-breve-vxnl}), rewritten as follows,
\begin{equation}
\label{eq-vx-kli-shell}
\breve{v}_{\text{x}}^{\text{KLI-HF}}(r) =\sum_{n'=1}^{n_\text{occ}}
 \frac{\rho_{n'}^{\text{HF}}(r)}{\rho^{\text{HF}}(r)} v_{\text{x}}^{(n')}(r)\,  ,
\end{equation}
is given for  $r \in S_{n}$ not only by  the respective shell potential $\tilde{v}_{\text{x}}^{(n)}(r)$,
but it also expressed there by the potentials $\tilde{v}_{\text{x}}^{(n')}(r)$ which correspond
to other shells ($n' \neq n$) and can be calculated for any $r$ with Eq. (\ref{vx-shell}).
The fact that, despite its significantly different structure, the potential
$\breve{v}_{\text{x}}^{\text{KLI-HF}}(r)$ (calculated with appropriate constants $\{C_{nl}\}$,  e.g., $\{C_{nl}^{\text{KLI-HF}}\}$)
is very close to 
 $v_{\text{x}}^{\text{pw,0}}(r)$ and $v_{\text{x}}^{\text{pw}}(r)$, and
 ultimately also to  $v_{\text{x}}^{\text{OEP}}(r)$, has been shown above to
 result from the relation (\ref{eq-hf-rel})  which holds for all $nl \in \{\text{occ}\}$
 within the occupied shells; for large $r$  the three approximate potentials and  the exact one
 nearly coincide with each other  due to  Eq. (\ref{eq-hf-rel-asymp}).

Finally, let us note that the potentials $v_{\text{x}}^{\text{pw,0}}(r)$ and
$v_{\text{x}}^{\text{pw}}(r)$ are identical for the Be atom
since, in this case, there are only two occupied orbitals, one in each of the two shells.

\subsection{Energy shifts. Step-like structure in the response part of exchange potential }
\label{sec-results-energy-shifts}
It is known that
the energies $\epsilon^{\text{KS}}_{nl}$ of the KS-OEP occupied atomic orbitals
 are higher than the corresponding
HF energies $\epsilon^{\text{HF}}_{nl}$
(except for the  HOMO energies which are nearly equal in the two schemes);
see Table  \ref{tab-dnl}.
The differences $\Delta \epsilon_{nl} = \epsilon^{\text{KS}}_{nl}-\epsilon^{\text{HF}}_{nl}$
are non-negative and, for given $l$, they are the larger the lower
shell index $n$ is.
The present results shed some light on these numerical findings as it is shown below.

Since the KS-OEP shifted orbital exchange  potentials $\tilde{v}_{\text{x;}nl}(r)$ and $\tilde{v}_{\text{x;}n+1,l}(r)$
(as well as the respective HF potentials)
match quite closely  at $r=r_{n,n+1}$
(cf. Figs. \ref{fig-Be_vx_vxa}, \ref{fig-Ar_vx_vxa}, \ref{fig-Zn_vx_vxa}),  we find
\begin{eqnarray}
\label{eq-diff-daa}
\Delta D_{nl,nl} \equiv D_{nl,nl}  - D_{n+1,l;n+1,l} \approx  v_{\text{x;}n+1,l}(r_{n,n+1})-v_{\text{x;}nl}(r_{n,n+1}) > 0 \; .
\end{eqnarray}
The latter  inequality results from the mathematical structure of
the Fock operator $\hat{v}_{\text{x}}^{\text{F}}({\bf r})$,
Eqs. (\ref{eq-vxfock-ylm}), (\ref{eq-fxnl-gen}), (\ref{eq-fxnl-vl}) 
presumably because
the orbital $\chi_{nl}(r)$ is localized
closer to the nucleus than $\chi_{n+1,l}(r)$.
This argument is certainly valid for the Be atom.
In this case,
the terms proportional to $v_0(1s,2s;r)$, which are  present in  Eqs. (\ref{eq-v1s-be}), (\ref{eq-v2s-be}),
are negligible  at the point $r=r_{12}=r_{12}^{\text{HF}}=0.954\, \text{a.u.}$
 where $\chi_{2s}(r)/\chi_{1s}(r)=0.69$ (see Fig. \ref{fig_Be_vxnl_part});
as a result, we have
 \begin{eqnarray}
 \label{eq-v2s-minus-v1s}
v_{\text{x;}2s}(r_{12})-v_{\text{x;}1s}(r_{12}) \approx    v_0(2s,2s;r_{12}) - v_0(1s,1s;r_{12}) \, .
 \end{eqnarray}
 The latter difference can be found by integrating the  equation  (\ref{eq-v1s2s-deriv})
 (here for the terms $v_0(nl,nl;r)$ defined with the KS orbitals) ,
 \begin{eqnarray}
 \label{eq-v2s0-minus-v1s0}
 v_0(2s,2s;r_{12}) - v_0(1s,1s;r_{12}) =
 \int_{\infty}^{r_{12}} dr'  \frac{1}{(r')^2} \left(  Q_{1s}(r') -  Q_{2s}(r') \right)  >0  \, ,
 \end{eqnarray}
 and it is positive since  the relation  $Q_{1s}(r) > Q_{2s}(r)$ holds for any $r$.
Let us note here that the approximate relation (\ref{eq-diff-daa})
is not satisfied very tightly for the closed-$l$-shell atoms other than Be
since
the differences $\Delta  v_{\text{x;}n+1,l}(r) \equiv v_{\text{x;}n+1,l}(r)-v_{\text{x;}nl}(r)$
change quite rapidly around $r=r_{n,n+1}$
(due to very  different slopes of the  orbital exchange potentials from the neighboring shells;
see Figs. \ref{fig-Be_vx_vxa}(e), \ref{fig-Ar_vx_vxa}, \ref{fig-Zn_vx_vxa})
while the point $r$ where
the potentials $\tilde{v}_{\text{x;}nl}(r)$ and $\tilde{v}_{\text{x;}n+1,l}(r)$ intersect
slightly differs (except for the Be atom) from the shell border $r_{n,n+1}=r_{n,n+1}^{\text{HF}}$
(defined in Sec. \ref{sec-results-represent-vx-shell-resolved}); cf. Fig. \ref{fig-Ar_vx_vxa}(d).
%
However, as it is seen in Table \ref{tab-dnl},  the differences $\Delta  v_{\text{x;}n+1,l}(r_{n,n+1})$
have quite similar value
and definitely the same sign as  the corresponding constants $\Delta D_{nl,nl}$.

Further, we can express $D_{nl,nl}$ as follows
\begin{equation}
\label{eq-daa-sumdiff}
D_{nl,nl} = D_{\tilde{n}(l)l,\tilde{n}(l)l} +
\sum_{n'=n}^{\tilde{n}(l)-1 } \Delta D_{n'l,n'l}
\end{equation}
for $n<\tilde{n}(l)$ where the symbol $\tilde{n}(l)$  denotes
the largest shell index $n$ among the KS-OEP occupied
orbitals $\chi_{nl}(r)$ with given orbital number $l$.
Thus, according to Eq. (\ref{eq-daa-sumdiff}) and the inequality
(\ref{eq-diff-daa}), the energy shift $\Delta \epsilon_{nl} \approx D_{nl,nl}$ grows with
decreasing $n$ and, consequently,  it is positive for $n<\tilde{n}(l)$ provided
the shift $D_{\tilde{n}(l)l,\tilde{n}(l)l}$ is non-negative.
The latter condition is satisfied
by the HOMO shift $D_{H,H}$ which vanishes.
For other orbitals $\chi_{\tilde{n}(l)l}$,
the   relation $D_{\tilde{n}(l)l,\tilde{n}(l)l}>0$ is established numerically
but understanding its origin needs further study.


The revealed representation of the exact exchange potential $v_{\text{x}}^{\text{OEP}}(r)$
with the (both HF and KS) orbital (or shell) exchange potentials does {\em not} result from  the
characteristic properties of
its response part
\begin{equation}
\label{eq-vx-resp}
v_{\text{x}}^{\text{resp}}(r)=v_{\text{x}}^{\text{OEP}}(r)-v_{\text{x}}^{\text{Sl}}(r)=
 v_{\text{x}}^{\text{ES}}(r)+v_{\text{x}}^{\text{OS}}(r) \, .
\end{equation}
This term has been found numerically \cite{leeuwen94} to have
a nearly step-like
dependence on $r$ where each step corresponds to an atomic shell.
The main part of   $v_{\text{x}}^{\text{resp}}(r)$ is the energy-shift (ES) term
\begin{equation}
\label{eq-vx-es-rad-orb}
v_{\text{x}}^{\text{ES}}(r)=
\frac{\sum_{nl}^{\text{occ}} (2 l +1) D_{nl,nl}\chi_{nl}^2(r)}{\sum_{nl}^{\text{occ}} (2 l +1) \chi_{nl}^2(r)}
\end{equation}
obtained from Eq. (\ref{eq-vx-es}).
The step-like $r$-dependence of
$v_{\text{x}}^{\text{resp}}(r) \approx v_{\text{x}}^{\text{ES}}(r)$
is briefly explained  in Ref. \onlinecite{leeuwen95} by noting that
within a given shell $S_n$ the orbitals $\chi_{n'l'}(r)$, $n' \neq n$, corresponding to other shells,
are small so that they can be neglected in Eq. (\ref{eq-vx-es-rad-orb}).
This argument can be supplemented  by
the numerical fact that
the different occupied orbitals  $\chi_{nl}(r)$ ($l \in {\cal L}_n$) from the $n$-th shell
have similar shapes and magnitudes within the respective shell region $S_n$.

\section{CONCLUSIONS}

In summary,
we find that when, for each HF orbital,
a suitably chosen (orbital-specific)
constant  shift is added to the Fock exchange operator in the HF equation,
the electrons occupying different
HF orbitals are subject to very similar local exchange potentials (as well as the total ones)
within the atomic regions
where the radial probability densities of the respective orbitals are substantial.
This proximity is particularly tight
for the shifted exchange potentials of the orbitals that belong to the same shell
and  it holds in the region of this shell.
Thus, the occupied HF orbitals are only very slightly disturbed when the
orbital-specific shifted exchange potentials are replaced in the HF equation
with a common exchange  potential  that lies very close to them within their respective shell regions;
simultaneously, the corresponding orbital energies change considerably
since  the applied shifts are quite sizeable.
As a result, the DFT exact exchange potential $v_{\text{x}}^{\text{OEP}}(r)$
(obtained in the OEP approach by minimizing the HF-like total energy expressed in terms
of the KS orbitals coming from a common local total potential)
is  very well represented in each shell  with the HF shifted orbital exchange potentials
from this shell, and, even slightly better, with their weighted average
-- the shell exchange potential, Eq. (\ref{vx-shell}).
Thus, the shape of the DFT  exchange potential in atoms,
as well as  its strongly shell-dependent slope, are, in fact, determined
by the $r$-dependences of the individual HF orbital exchange potentials within their corresponding
shells.

The revealed  properties of the shifted orbital exchange potentials result from
the more general relation  (\ref{eq-hf-rel}) satisfied 
by the Fock exchange operator  and the HF orbitals.
Thus, it is in fact this relation that explains
the outstanding proximity of the HF and KS orbitals in the closed-$l$-shell atoms
as well as  the high-quality of the KLI and  LHF(CEDA) approximations
to the exact exchange potential $v_{\text{x}}^{\text{OEP}}$.
However, since  these approximations
are expressed in terms of the exchange potentials
of {\em all} occupied orbitals (at a given point $r$),
one of  {\em qualitatively} new achievements of  the present work is showing
that the potential $v_{\text{x}}^{\text{OEP}}(r)$ can be represented,
with equally high accuracy, by the (HF or KS) {\em individual} shifted orbital exchange potentials
within their corresponding shells.
An intermediate stage between these  two types of  representation is obtained with
the piecewise function  formed  with the shell exchange potentials.
It is also shown that the positive values of the differences
$\epsilon_{nl}^{\text{KS}}-\epsilon_{nl}^{\text{HF}}$ between the energies
of the respective KS and HF orbitals, as well as their increase with decreasing $n$
are related to the  differences between the orbital exchange potentials
from neighboring shells at the shell borders.
Finally, it should be stressed that the presently obtained shell-resolved mapping between the HF orbital
exchange potentials and the DFT exact exchange potential is {\em not}
related to the previously established step-like structure of the response part of
the exchange potential.

\appendix*
\section{Asymptotic dependence of Hartree-Fock exchange orbital potentials}

In the atomic region outside the occupied shells,
the HF orbitals have the following asymptotic dependence \cite{HMS69,HSS80,DJM84,IO92}
\begin{equation}
\label{eq-chi-hf-asymp}
\chi_{nl}^{\text{HF}}(r) \sim \left( r^{\alpha_{nl}}+ b_{nl} r^{\alpha_{nl}-1}\right) e^{-\beta_H r}
\end{equation}
where the coefficient $\beta_H=\sqrt{-2\epsilon_{H}^{\text{HF}}}$ is common for all $nl$ while
the constants $b_{nl}$ and the powers $\alpha_{nl}$ are orbital-specific.
The largest $\alpha_{nl}$  is found for  the HOMO and it is equal to $\alpha_{H}=1/\beta_H$
for neutral atoms. For other HF orbitals the powers $\alpha_{nl}$ depend on the orbital number $l$,
i.e.,
\begin{subequations}
\label{eq-chi-hf-asymp2}
\begin{eqnarray}
  \alpha_{nl} &=& \alpha_{H} - 3 \hspace{3.0cm} (l=l_H \neq 0 \,, \ nl\neq H),   \\
  \alpha_{nl} &=& \alpha_{H} -2(l_{\text{min}}+1) \hspace{1.4cm}  (l=l_H =0 \, , \ nl\neq H),\\
  \alpha_{nl} &=& \alpha_{H} - |l-l_H|  -1 \hspace{1.2cm}  (l\neq l_H ) ,
\end{eqnarray}
\end{subequations}
Here, $l_H$ denotes the HOMO orbital number and $l_{\text{min}}$
is the smallest non-zero $l$ within the set of the occupied HF orbitals  in a given atom.
The above asymptotic dependence (\ref{eq-chi-hf-asymp}) is valid for all atoms other than Be.

In the asymptotic region, the HF hamiltonian $\hat{h}_{\text{HF}}({\bf r})$, Eq. (\ref{eq-hf}),
is dominated by the kinetic and the exchange terms  since, for a neutral atom,  the sum
$v_{\text{ext}}({\bf r})+v_{\text{es}}^{\text{HF}}({\bf r})$ decays exponentially
(as  $e^{-2\beta_H r}$) for large $r$.
Thus, the HF radial equation has following asymptotic form
\begin{equation}
\label{eq-hf-rad-asymp}
\left [ -\frac{1}{2} \frac{d^2}{dr^2}  + \frac{l(l+1)}{2r^2} +
v_{\text{x;}nl}^{\text{HF}} (r)\right ] \chi_{nl}^{\text{HF}} (r)=\epsilon_{nl}^{\text{HF}} \chi_{nl}^{\text{HF}} (r)
\hspace{1cm}  (\text{large } r) \ ,
\end{equation}
and, by dividing its both sides with $\chi_{nl}^{\text{HF}} (r)$, we obtain
 \begin{equation}
\label{eq-vxnl-hf-asymp2}
v_{\text{x;}nl}^{\text{HF}} (r) = \epsilon_{nl}^{\text{HF}}+
\frac{d^2\chi_{nl}^{\text{HF}}(r)/dr^2}{2\chi_{nl}^{\text{HF}}(r) }    -
\frac{l(l+1)}{2r^2} + o\left(\frac{1}{r^2}\right) \, .
\end{equation}
When the asymptotic dependence (\ref{eq-chi-hf-asymp}) of the orbital $\chi_{nl}^{\text{HF}} (r)$
is applied, the general asymptotic form (\ref{eq-vxnl-hf-asymp})
of the HF exchange orbital potentials $v_{\text{x;}nl}^{\text{HF}} (r)$ is found.

By using the explicit expression for $v_{\text{x;}nl}^{\text{HF}} (r)$
(given by Eqs. (\ref{eq-fxnl-gen}), (\ref{eq-fxnl-vl}), (\ref{eq-vxnl}) with the HF orbitals),
one readily finds the asymptotically dominating term $(-1/r)v_0(H,H;r=\infty)=-1/r$
in the HOMO exchange potential $v_{\text{x;}H}^{\text{HF}} (r)$; cf. Eq. (\ref{eq-hf-rel-asymp}).
The same term, $(-1/r)v_0(nl,nl;r=\infty)=-1/r$, is present in the asymptotic dependence of
any  potential $v_{\text{x;}nl}^{\text{HF}} (r)$, but, for $nl\neq H$,  it also includes
other terms  
which are proportional
to $1/r$ or tend to constant values for  $r\rightarrow \infty$ (the latter
contribute to the constant term $-\left(\epsilon_{H}^{\text{HF}}-\epsilon_{nl}^{\text{HF}}\right)$
in Eq. (\ref{eq-vxnl-hf-asymp})).
For instance, the potential $v_{\text{x;}2p}^{\text{HF}} (r)$
contains the terms proportional to
$r^{-3}\chi_{3p}(r) v_{2}(3p,2p;r) / \chi_{2p}(r)$
and $r^{-2}\chi_{3s}(r) v_{1}(3s,2p;r) / \chi_{2p}(r)$
which depend like  $(c_1 +c_2/r)$ and $c_3/r$, respectively, for large $r$;
here $c_1$, $c_2$, and $c_3$ are constants; these asymptotic dependences can be derived using
Eqs. (\ref{eq-chi-hf-asymp}) and (\ref{eq-chi-hf-asymp2})

For the Be atom,  the  two occupied $s$ orbitals decay as
$\chi_{n0}^{\text{HF}}(r) \sim r^{1/\beta} e^{-\beta r}$ ($n=1,2$) where
$\beta=\sqrt{-2\epsilon_{n0}^{\text{HF}}}$.
Thus, according to Eq. (\ref{eq-vxnl-hf-asymp2}), the potentials  $v_{\text{x;}n0}^{\text{HF}} (r)$ ($n=1,2$)
vanish at $r\rightarrow \infty$. They have the  same asymptotic dependence $(-1/r)v_0(n0,n0;r=\infty)=-1/r$,
which results  from Eqs. (\ref{eq-v1s-hf-be}),  (\ref{eq-v2s-hf-be}) and the definition of  the functions
$v_{l''}^{\text{HF}}(n'l',nl;r)$, Eq. (\ref{eq-fxnl-vl}).

\acknowledgments
Discussions with A. Holas are gratefully acknowledged.

\pagebreak
\newcolumntype{e}[1]{D{.}{.}{#1}}
\begin{table*}[h]
\caption{
\label{tab-dphi}
 The norms of $\Delta \phi_{nlm} =\phi_{nlm}^{\text{HF}}-\phi_{nlm}$,
 $\delta\phi_{nlm}$, $\Delta \phi_{nlm} - (-\delta\phi_{nlm})$ and
 (in the last column) the upper bound of
 $\|\delta\phi_{nlm}\|$, Eq. (\ref{eq-dphi-upper-bound}),
 for the occupied orbitals $\phi_a=\phi_{nlm}$ in the Be and Ar atoms;
 see text for details.   }
 \begin{ruledtabular}
\begin{tabular}{ccccce{8}}
  atom & orbital   &  $\| \Delta \phi_{nlm}\|$  & $\| \delta\phi_{nlm} \|$ &
 $\|  \Delta \phi_{nlm} - (-\delta\phi_{nlm}) \|$ &
   \multicolumn{1}{c}{$\|W_{nl}^{\perp\text{;rad}}\| / | \epsilon_{n+1,l}-\epsilon_{nl} |$}  \\
   & $(nl)$  &  $(\times 10^{-3})$ &  $(\times 10^{-3})$ & $(\times 10^{-3})$
   &  \multicolumn{1}{c}{$(\times 10^{-3})$}  \\
  \hline
 Be & $1s$ & 6.0890  &   6.6865 &    0.6253  &    11.2134 \\
      & $2s$ & 5.7655 &    6.3021 &    0.6416  &    242.9996 \\
 \hline
 Ar   & $1s$ &  1.2594 &  1.2752   &   0.0305 &  3.3526  \\
       & $2s$ &  6.2281 &   6.5057  &   0.2929 &  22.0419\\
       & $2p$ &  4.3019  &  4.5323  &   0.2467 &   82.4518  \\
       & $3s$ &  5.8187  &  6.4366  &   0.8003 &  122.1715   \\
       & $3p$ &  4.3474  &  4.5782  &   0.3428 &  242.2264  \\
\end{tabular}
\end{ruledtabular}
\end{table*}

\pagebreak

\begingroup
\squeezetable
\begin{table*}[h]
\caption{
\label{tab-dnl}
The  HF and exchange-only KS (OEP) orbital energies, $\epsilon_{nl}$, $\epsilon_{nl}^{\text{HF}}$,
in the Ar atom. The difference $\epsilon_{nl}-\epsilon_{nl}^{\text{HF}}$
compared with the constant shifts $D_{nl}^{\text{OEP}}\equiv D_{nl}[v_{\text{x}}^{\text{OEP}}]=\delta \epsilon_{nl}$ and
$\{C_{nl}^{\text{KLI-HF}}\}=D_{nl}^{\text{HF}}[v_{\text{x}}^{\text{KLI-HF}}]$,
obtained with the OEP and KLI-HF exchange potentials, respectively.
The constants
$\Delta D_{n,n+1}^{(l)}=D_{nl}^{\text{OEP}}-D_{n+1,l}^{\text{OEP}}$ compared to
the differences $\Delta v_{\text{x;}n,n+1}^{(l)}(r)= v_{\text{x;}n+1,l}(r)-v_{\text{x;}nl}(r)$
of the KS-OEP orbital exchange potentials at the shell borders
$r=r_{n,n+1}^{\text{HF}}$. See text for details.
Note that if the point $r_{12}^{\text{HF}}$ was moved
by just  $0.01\, \text{a.u.}$, to $r=0.137\, \text{a.u.}$ for the Ar atom,
the considerably modified value $\Delta v_{12}^{(0)}(r) = 2.98\, \text{hartree}$ would be obtained  .
The HF orbitals and their energies used in the calculations are taken from Ref. \onlinecite{bunge93}.
All energies and radii are given in hartrees.
}
 \begin{ruledtabular}
\begin{tabular}{cc e{8}e{14}e{9}e{14}e{7}ddd}
  \multicolumn{1}{c}{atom} & $\; nl \; $ &  \multicolumn{1}{c}{$\epsilon_{nl}^{\text{HF}}$} &
  \multicolumn{1}{c}{$\epsilon_{nl}$} &
   \multicolumn{1}{c}{$\epsilon_{nl}-\epsilon_{nl}^{\text{HF}}$} &
   \multicolumn{1}{c}{$D_{nl}^{\text{OEP}}$} &  \multicolumn{1}{c}{$C_{nl}^{\text{KLI-HF}}$} &
 \multicolumn{1}{c}{$\Delta D_{n,n+1}^{(l)}$} &
 \multicolumn{1}{c}{$\Delta v_{\text{x;}n,n+1}^{(l)}$} & \multicolumn{1}{c}{$r_{n,n+1}^{\text{HF}}$} \\
 \colrule

  Be & $1s$ & -4.732\;669 & -4.125\; 699\; 368\;  4 & 0.606\; 969  & 0.606\; 401\;428\;6 & 0.562\;484 & 0.606   & 0.607 & 0.954 \\
       & $2s$ & -0.309\; 269 & -0.309\; 227\; 738\;  5   & 0.000\; 041  & 0.0                & 0.0         &   &  &  \\[2mm]

 \colrule

 Ar   & $1s$  & -118.610\;349 & -114.452\;154\;608\;6   &  4.158\;194  &  4.156\;319\;209\;3 & 4.153\;224  & 2.991 & 4.426 & 0.127 \\
       & $2s$  &   -12.322\;152 &  -11.153\;224\;215\;2   &  1.168\;928  &  1.165\;666\;206\;9 &  1.126\;130  & 0.988 & 1.031  & 0.729 \\
       & $2p$  &    -9.571\;464  &   -8.733\;757\;145\;4   &   0.837\;707 &   0.837\;222\;865\;2 &  0.764\;760  & 0.837 & 0.963 & 0.729\\
       & $3s$  &    -1.277\;352   &   -1.099\;246\;843\;1 &   0.178\;105 &    0.178\;063\;552\;5 &  0.180\;419  &             &          &  \\
       & $3p$  &    -0.591\;016  &    -0.590\;751\;487\;8   &   0.000\;265 &  0.0                &  0.0            &             &          & \\

\end{tabular}
\end{ruledtabular}
\end{table*}
\endgroup

\pagebreak

\noindent
FIGURE CAPTIONS\\

\noindent
Fig. 1.
OS norm square  $\|\delta \phi_a\|^2$ (grey bars)
and the contributions $c_{n'l;nl}^2$ (stacked bars) to it from bound states
$\phi_{n'lm}$, for the occupied  states $\phi_{nlm}$ in the Ar atom;
the contributions from the occupied states
are marked with the hatch patterns; the $1s$ bars are magnified by the factor 20.
The results are obtained in the exchange-only KS-OEP scheme.  \\

\noindent
Fig. 2.
(a) KS-OEP radial electron density $\rho$ (per  spin) and
(b,c) the term $F_{\text{x;}nl}+D_{nl,nl}\chi_{nl}$ (dashed and dotted lines) compared to
$v_{\text{x}}^{\text{OEP}}\chi_{nl}$ (solid lines)  in the Be atom, $(nl)=1s,2s$.
(d,e) The potentials
$v_{\text{x}}^{\text{OEP}}$ (solid line),
$v_{\text{x;}1s}$ (dashed-dotted line),
$\tilde{v}_{\text{x;}1s}$ (dotted line),
$v_{\text{x;}2s}=\tilde{v}_{\text{x;}2s}$ (dashed line),
$v_{\text{x}}^{\text{KLI-HF}}$ (long-dashed line in the insert (e)).
The HF radial electron density $\rho^{\text{HF}}$ and
the HF potentials $v_{\text{x;}nl}^{\text{HF}}$, $\tilde{v}_{\text{x;}nl}^{\text{HF}}$,
$nl=1s,2s$, follow  $\rho$, $v_{\text{x;}nl}$ and $\tilde{v}_{\text{x;}nl}$, correspondingly,
within the figure resolution. The up and down arrows mark the points
$r_{12}^{\text{HF}}$ and $r_{1}^{\text{min}}$, respectively.\\

\noindent
Fig. 3.
(a) KS-OEP radial electron density $\rho$ (per spin) and
(b,c) the potentials $v_{\text{x}}^{\text{OEP}}$ (solid
line), $\tilde{v}_{\text{x;}nl}$ (dashed and dotted lines) in the Ar atom.
(d) The differences
$\Delta v_{\text{x;}nl}=  v_{\text{x;}nl}-v_{\text{x}}^{\text{OEP}}$ (dashed  lines)
and $\Delta \tilde{v}_{\text{x;}nl}=\tilde{v}_{\text{x;}nl}-v_{\text{x}}^{\text{OEP}}$ (solid lines),
each shown within the $r$-interval including the corresponding shell  $S_n$ and slightly overlaping
the neighboring shells ($S_{n-1}$ and/or $S_{n+1}$).
The HF radial electron density $\rho^{\text{HF}}$ and the HF potentials
$\tilde{v}_{\text{x;}nl}^{\text{HF}}$ as well as the differences
$\Delta v_{\text{x;}nl}^{\text{HF}}=v_{\text{x;}nl}^{\text{HF}}-v_{\text{x}}^{\text{OEP}}$ and
$\Delta \tilde{v}_{\text{x;}nl}^{\text{HF}}=\tilde{v}_{\text{x;}nl}^{\text{HF}}-v_{\text{x}}^{\text{OEP}}$
follow $\rho$, $\tilde{v}_{\text{x;}nl}$, $\Delta v_{\text{x;}nl}$, and $\Delta \tilde{v}_{\text{x;}nl}$, correspondingly,
 within the resolution of the respective figure.
The up and down arrows mark the points
$r_{n,n+1}^{\text{HF}}$ and $r_{n}^{\text{min}}$, respectively.\\

\noindent
Fig. 4.
Results for the Zn atom; the description of the panels (a)-(d) as in Fig. \ref{fig-Ar_vx_vxa}.
The HF quantities $\rho^{\text{HF}}$,
$\tilde{v}_{\text{x;}nl}^{\text{HF}}$,
$\Delta v_{\text{x;}nl}^{\text{HF}}$, $\Delta \tilde{v}_{\text{x;}nl}^{\text{HF}}$
follow $\rho$, $\tilde{v}_{\text{x;}nl}$, $\Delta v_{\text{x;}nl}$, and $\Delta \tilde{v}_{\text{x;}nl}$, correspondingly,
 within the resolution of the respective figure.\\

 \pagebreak

\noindent
Fig. 5.
(a,b,c) Asymptotic dependence of the potentials 
$v_{\text{x;}nl}^{\text{HF}}$  (solid lines)  and $v_{\text{x;}nl}$  (dotted lines)
compared with the HF asymptotic limits, equal to $-\left(\epsilon_{H}^{\text{HF}}-\epsilon_{nl}^{\text{HF}}\right)$
 (horizontal dashed lines) , Eq. (\ref{eq-vxnl-hf-asymp}), in the Ar atom.
(c) The HOMO exchange potentials
$v_{\text{x;}3p}^{\text{HF}}$ and $v_{\text{x;}3p}$ (which follow each other within the figure resolution)
are compared with the $-1/r$ (dashed line) asymptotic dependence of $v_{\text{x}}^{\text{OEP}}$.
The results are obtained with the KS-OEP and HF orbitals calculated, with high accuracy,
by using  the pseudospectral method  \cite{CH07, MC09-prep}.
 Note that the divergence of $v_{\text{x;}1s}^{\text{HF}}(r)$ seen in the panel (a) results from
the node of the HF orbital $\chi_{1s}^{\text{HF}}(r)$ at $r=1.09 \,\text{a.u.} $; this node is also present in
$\chi_{1s}^{\text{HF}}(r)$ calculated with
the Slater-type-orbital expansion given in Ref. \onlinecite{bunge93}.\\

\noindent
Fig. 6.
Differences between approximate and exact exchange potentials:
 (a) $v_{\text{x}}^{\text{pw,0}}-v_{\text{x}}^{\text{OEP}}$,
 (b) $v_{\text{x}}^{\text{pw}}-v_{\text{x}}^{\text{OEP}}$,
 (c) $v_{\text{x}}^{\text{KLI-HF}}-v_{\text{x}}^{\text{OEP}}$,
 (d) $\breve{v}_{\text{x}}^{\text{KLI}}(r) - v_{\text{x}}^{\text{OEP}}=-v_{\text{x}}^{\text{OS}}$
 (Eqs. (\ref{eq-vx-oep}), (\ref{eq-vx-kli-oep})); see text for details.
The dashed lines correspond  to  $v_{\text{x}}^{\text{OEP}}-v_{\text{x}}^{\text{OEP}}=0$.
The up  arrows mark the points $r_{n,n+1}^{\text{HF}}$, $n=1,2$.\\

\noindent
Fig. 7.
 (a) KS-OEP orbital exchange potentials $v_{\text{x;}1s}$,  $v_{\text{x;}2s}$ (solid lines),
Eqs. (\ref{eq-v1s-be}), ( \ref{eq-v2s-be}),
 compared with
 the contributing functions $v_0(1s,1s;r)$,  $v_0(2s,2s;r)$, $v_0(1s,2s;r)=v_0(2s,1s;r)$ (dotted and dashed lines),
 Eq. (\ref{eq-fxnl-vl}),
 and (b) the ratios  $\chi_{2s}/\chi_{1s}$, $\chi_{1s}/\chi_{2s}$
 for the Be atom.
 The HF potentials $v_{\text{x;}n0}^{\text{HF}}\;$, Eqs. (\ref{eq-v1s-hf-be}), (\ref{eq-v2s-hf-be}),
 functions $v_0^{\text{HF}}(n0,n'0;r)$ and ratios
 $\chi_{n0}^{\text{HF}}/\chi_{n'0}^{\text{HF}}$ ($n,n'=1,2$) follow the corresponding KS-OEP quantities
 within the figure resolution. The up down arrow marks the difference
 $\Delta v_{\text{x;}n,n+1}^{(l)}(r)= v_{\text{x;}n+1,l}(r)-v_{\text{x;}nl}(r)$ at
 $r=r_{12}^{\text{HF}}=0.954 \,\text{a.u.}$; it is very close to $D_{1s,1s}^{\text{OEP}}=0.6064 \, \text{hartree}$;
 see Table \ref{tab-dnl}.

\pagebreak

\setcounter{figure}{0}
\begin{figure}[h]
\includegraphics*[width=8.5cm]{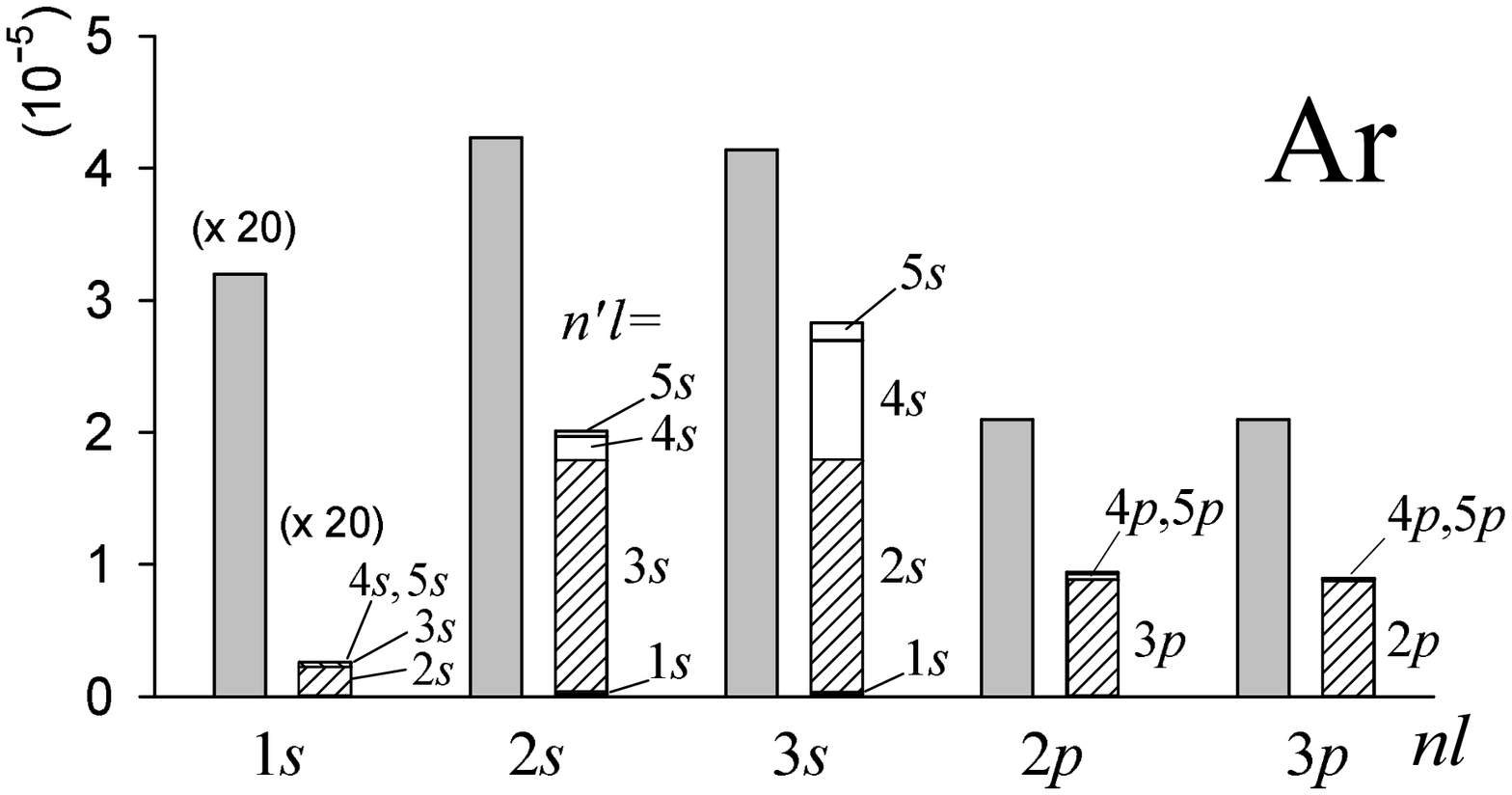}  
\caption{}
\label{fig-dchi-dist-Ar}
\end{figure}

\pagebreak

\begin{figure}[h]
\includegraphics*[width=8.5cm]{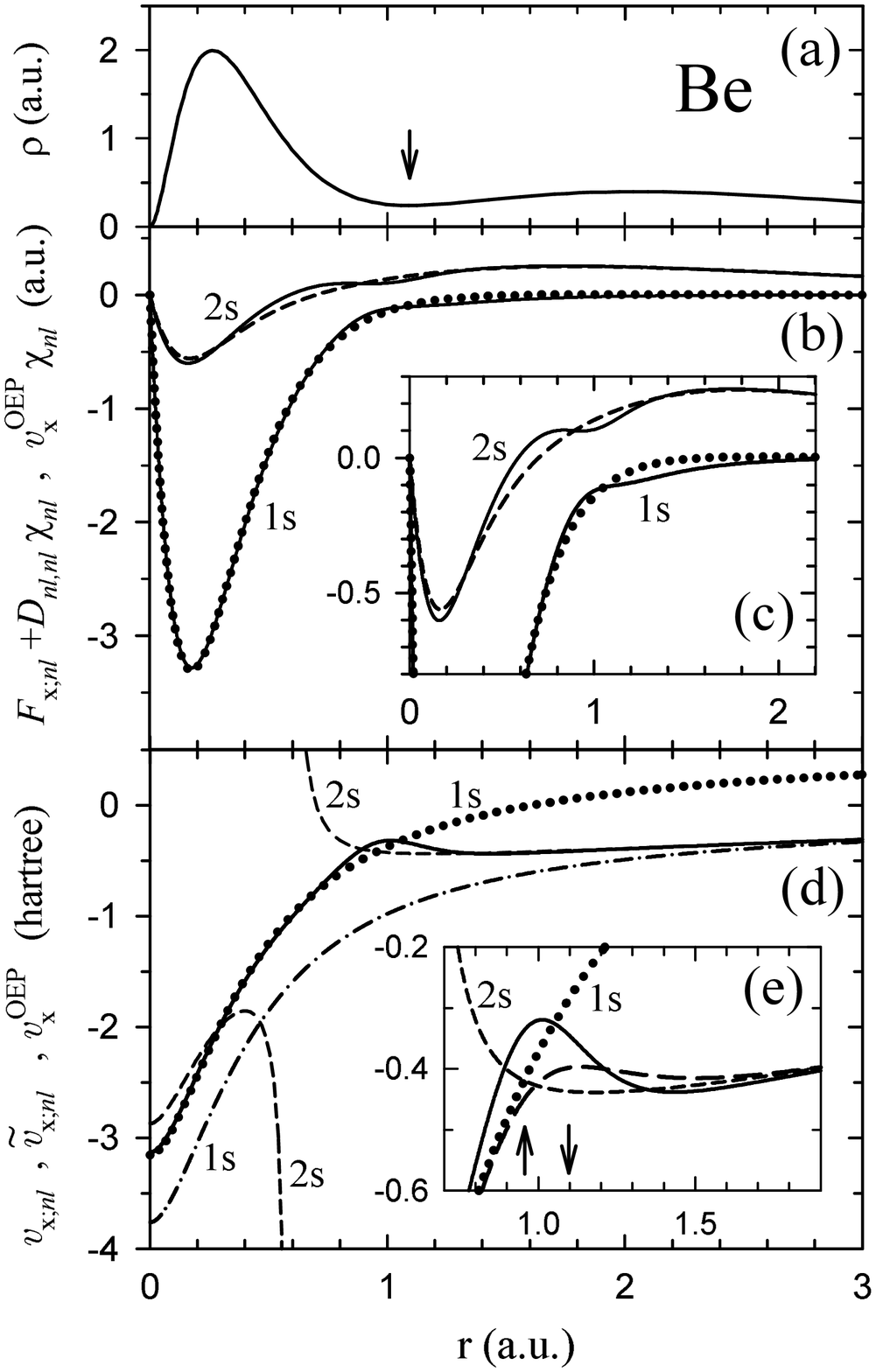}  
\caption{ }
 \label{fig-Be_vx_vxa}
\end{figure}

\pagebreak

\begin{figure}[h]
\includegraphics*[width=8.5cm]{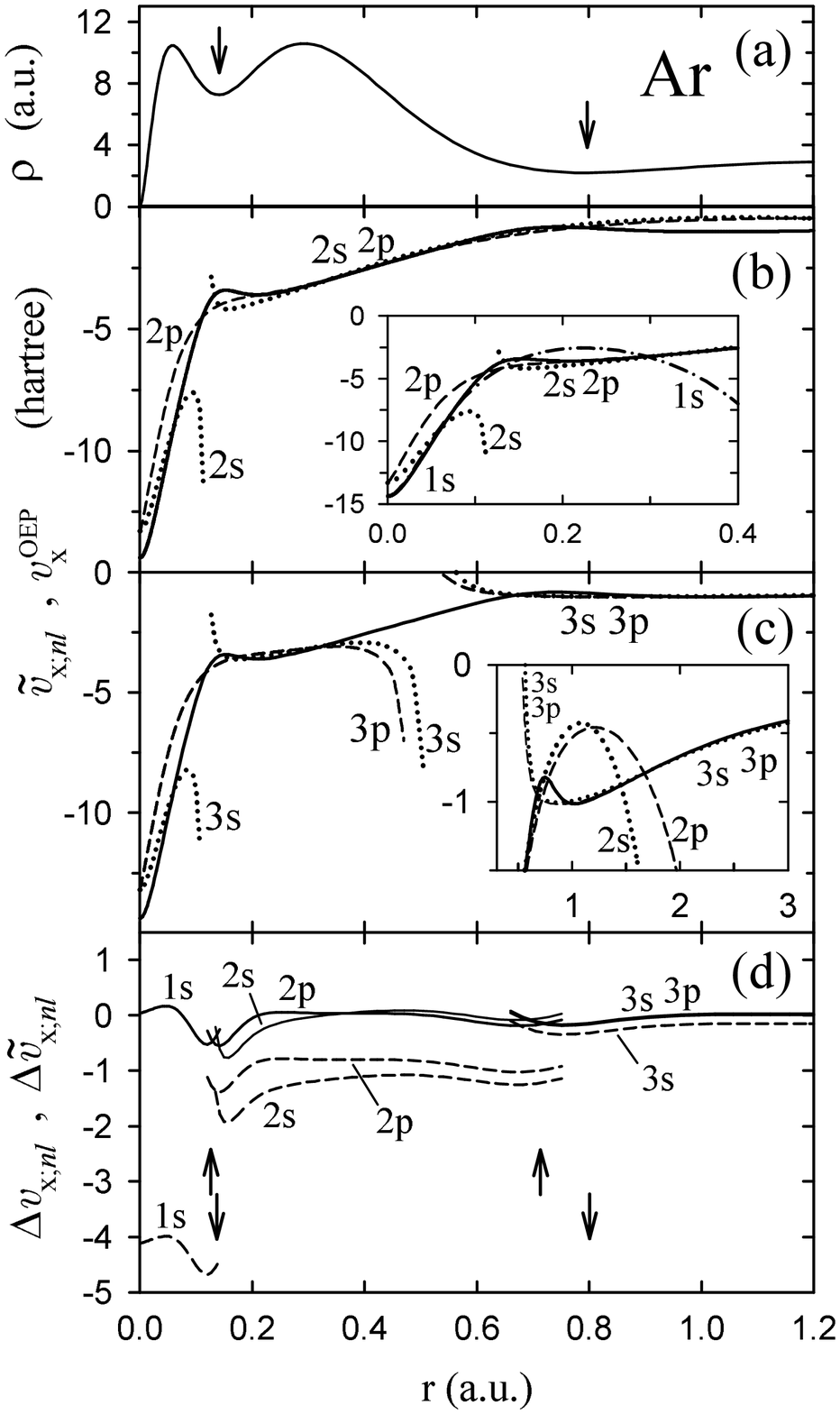}  
\caption{ }
\label{fig-Ar_vx_vxa}
\end{figure}

\pagebreak

\begin{figure}[h]
\includegraphics*[width=8.5cm]{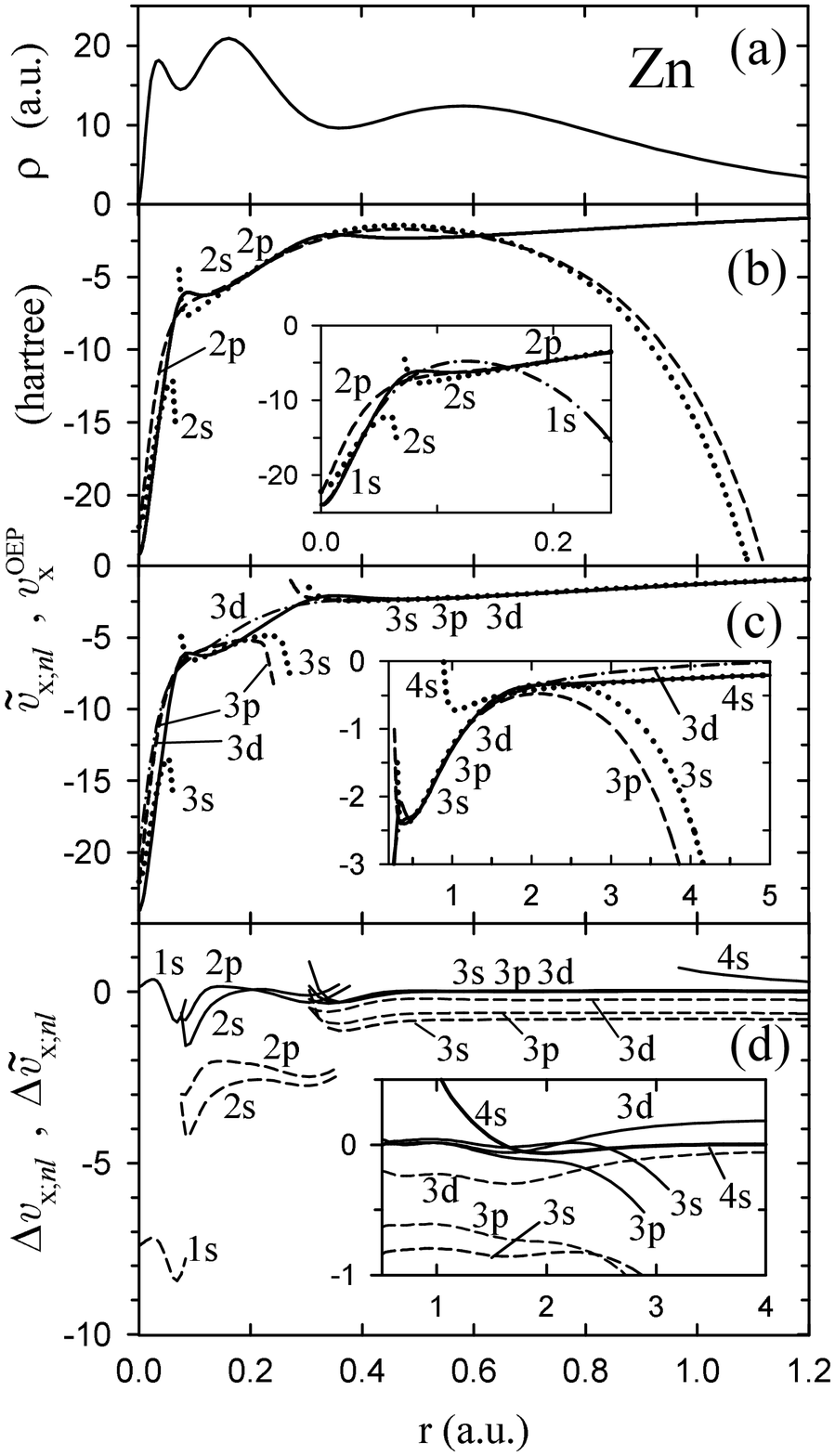}  
\caption{ }
\label{fig-Zn_vx_vxa}
\end{figure}

\pagebreak

\begin{figure}[h]
\includegraphics*[width=8.5 cm]{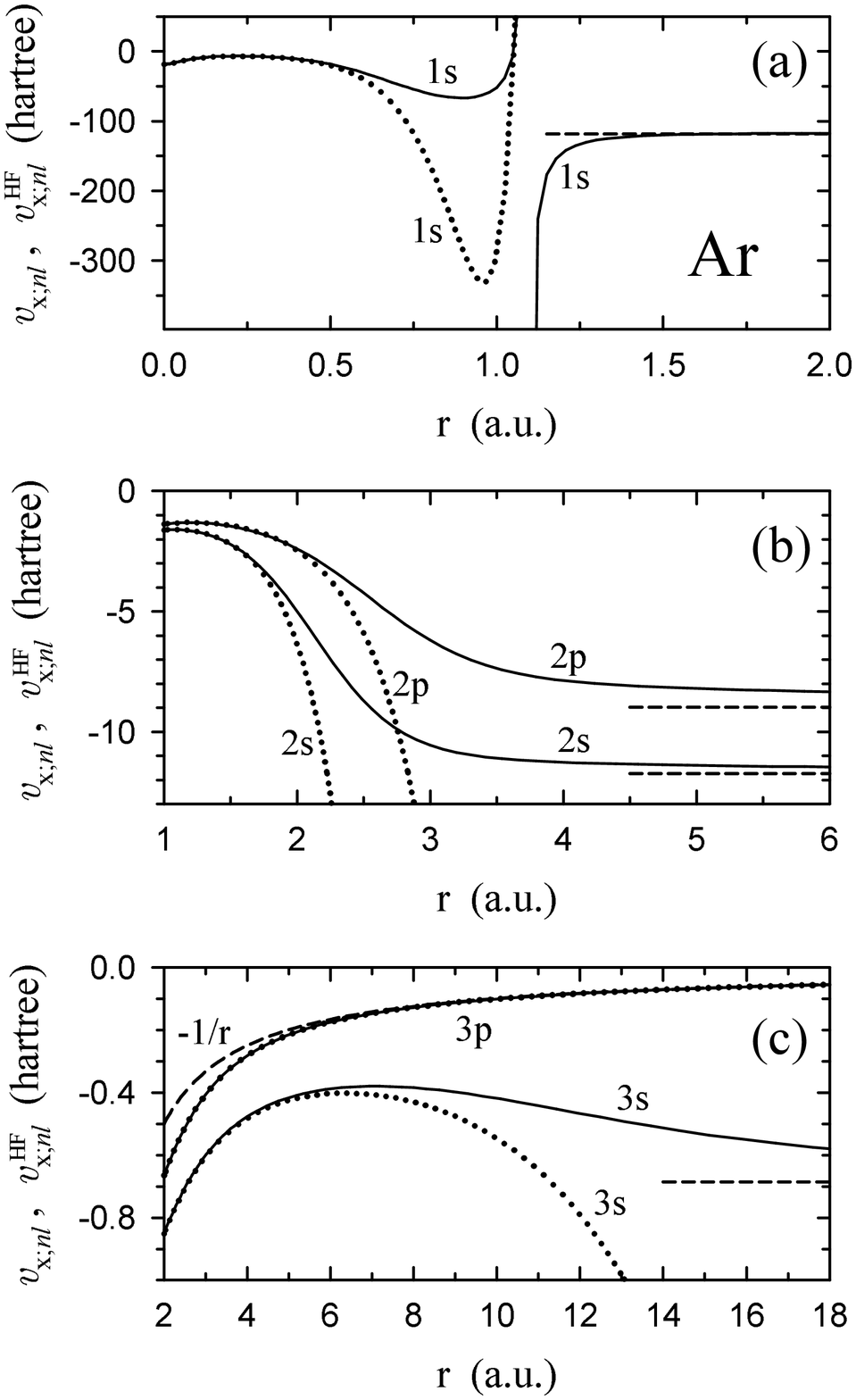}  
\caption{ }
 \label{fig-Be_Ar-asymp}
\end{figure}

\pagebreak

\begin{figure}[h]
\includegraphics*[width=8.5 cm]{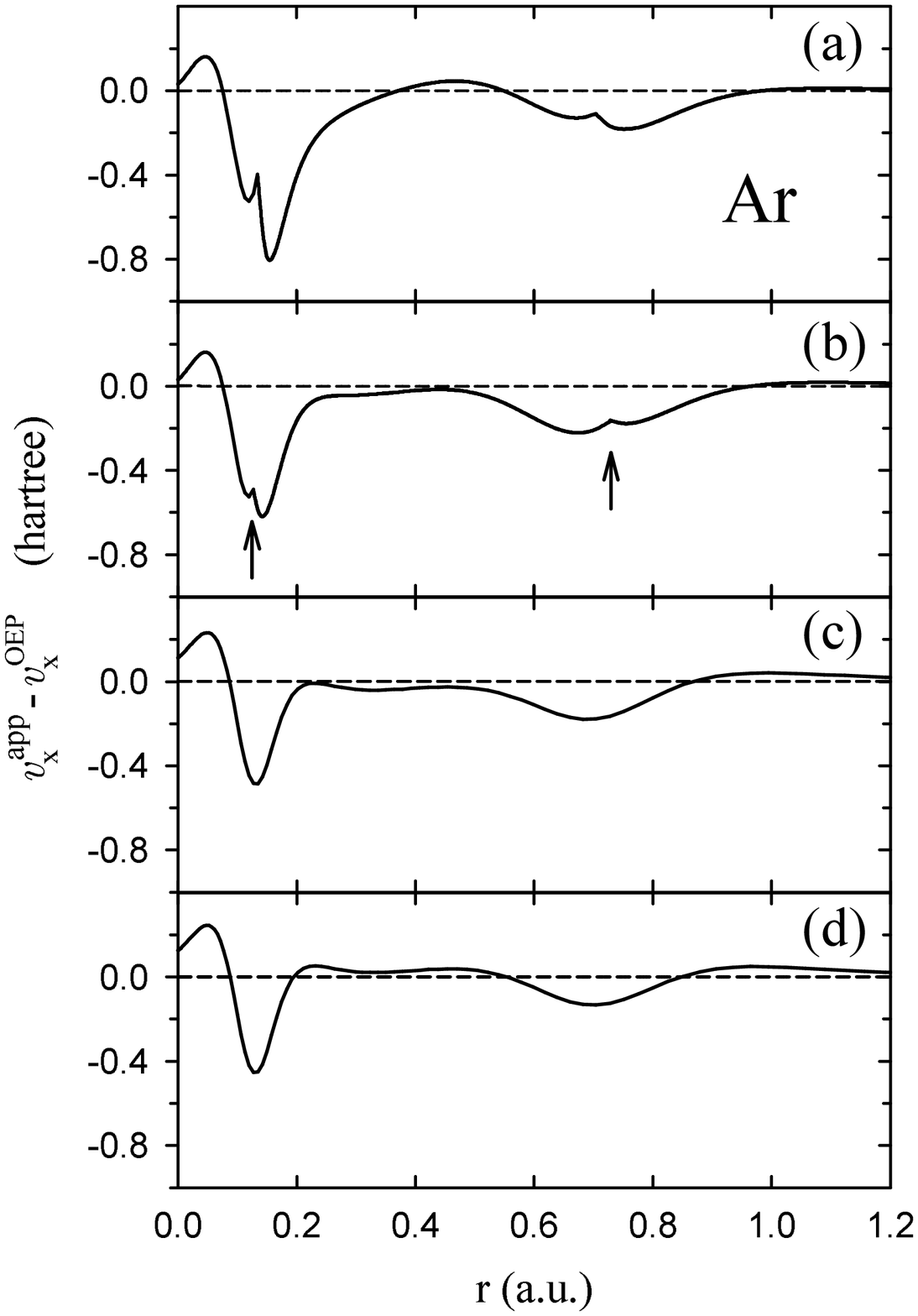}  
\caption{ }
\label{fig-Ar_vxapp-vxoep}
\end{figure}

\pagebreak

\begin{figure}[h]
\includegraphics*[width=8.5cm]{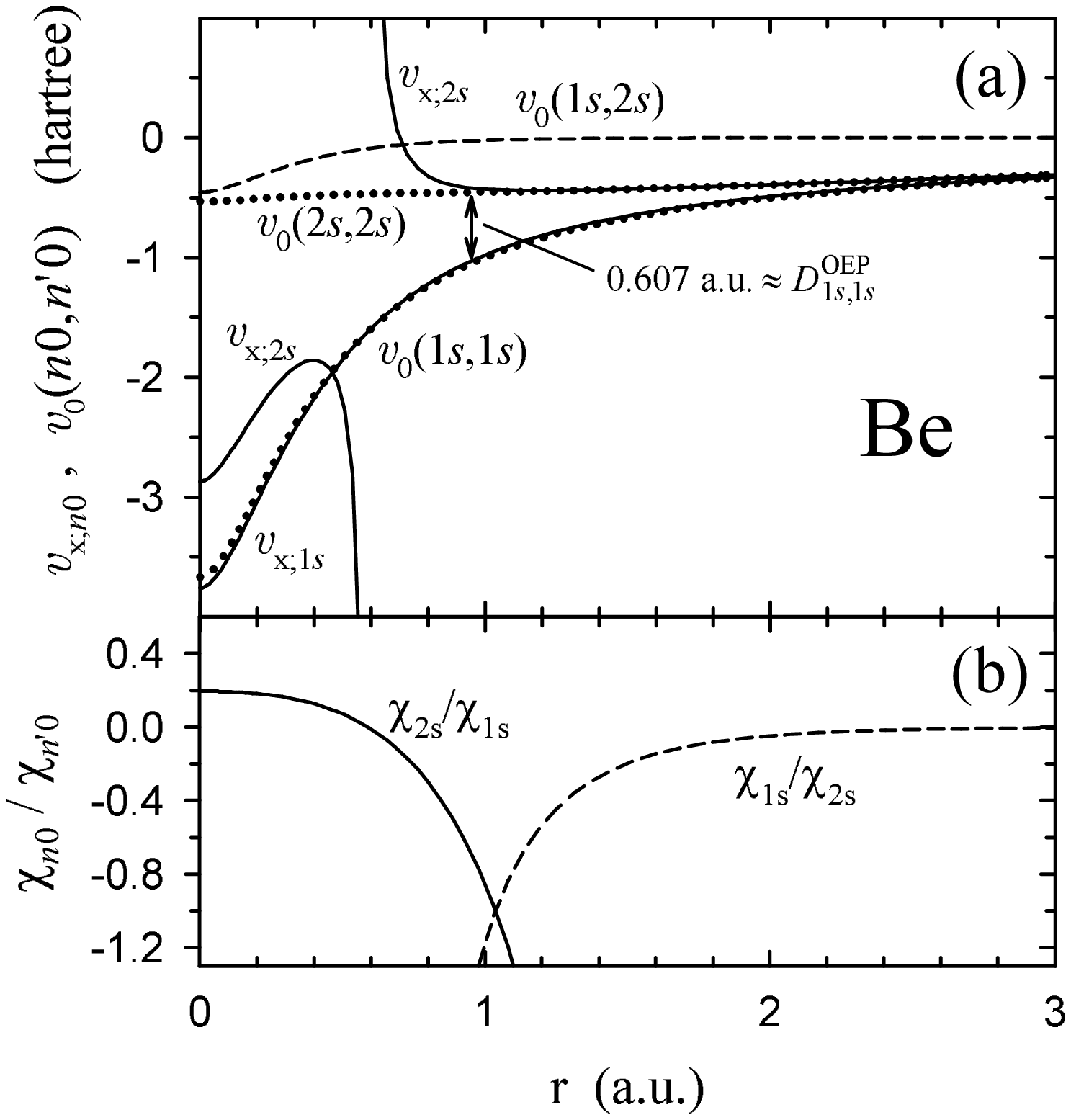}  
\caption{}
\label{fig_Be_vxnl_part}
\end{figure}


\begin{thebibliography}{99}
\bibitem{note-diff-en}
The latter is not valid  for $a=\text{HOMO}$, since $D_{aa}=0$,
but  $\Delta\epsilon_{a}=-4.13\cdot 10^{-5}\;\text{hartree}$ for Be  and
$-2.645\cdot 10^{-4}\;\text{hartree}$  for Ar; cf. Table \ref{tab-dnl}.
\bibitem{note-etot-vx}
The KS orbitals $\phi_a$ can be treated as  the functionals of the
exchange potential $v_{\text{x}}$ once they are found, in a selfconsistent way,
from the KS equation with the local potential $v_{\text{s}}=v_{\text{ext}}+v_{\text{es}}[\{\phi_a\}]+v_{\text{x}}$.
Similar functional dependence holds for a quantity, like the total
energy $E$, that depends on the orbitals.
The functional $E[v_{\text{x}}]=E[\{\phi_a[v_{\text{x}}]\}]$ is continuous because both functionals
$E[\{\phi_a\}]$ and $\phi_a[v_{\text{x}}]$ ($a=1,\ldots,N$) are continuous;
in the latter case the continuity is defined on the set of the physically unequivalent potentials
$v_{\text{x}}$, which satisfy the condition $v_{\text{x}}(r\rightarrow \infty)=0$
\end{thebibliography}
\end{document}